# Multi-bunch Feedback Systems

*M. Lonza, presented by H. Schmickler*
Elettra Synchrotron Light Laboratory, Sincrotrone Trieste S.C.p.A., Trieste, Italy

**Abstract**
Coupled-bunch instabilities excited by the interaction of the particle beam with its surroundings can seriously limit the performance of circular particle accelerators. These instabilities can be cured by the use of active feedback systems based on sensors capable of detecting the unwanted beam motion and actuators that apply the feedback correction to the beam. Advances in electronic technology now allow the implementation of feedback loops using programmable digital systems. Besides important advantages in terms of flexibility and reproducibility, digital systems open the way to the use of novel diagnostic tools and additional features. We first introduce coupled-bunch instabilities, analysing the equation of motion of charged particles and the different modes of oscillation of a multi-bunch beam, showing how they can be observed and measured. Different types of feedback systems will then be presented as examples of real implementations that belong to the history of multi-bunch feedback systems. The main components of a feedback system and the related issues will also be analysed. Finally, we shall focus on digital feedback systems, their characteristics, and features, as well as on how they can be concretely exploited for both the optimization of feedback performance and for beam dynamics studies.

## 1 Coupled-bunch instabilities

### 1.1 Introduction

The demand for high brightness in synchrotron light sources and luminosity in circular colliders require the storage of intense particle beams with many bunches. The interaction of these beams with the environment in which they move can give rise to collective effects called 'coupled-bunch instabilities'. Collective instabilities can be seen as follows: charged particles in a storage ring are subject to transverse (betatron) and longitudinal (synchrotron) oscillations that are normally damped by a natural damping mechanism. The electromagnetic field created by the bunches can interact with the surrounding metallic structures generating 'wake fields', which act back on the beam producing growth of the oscillations. If the growth is stronger than the damping the oscillation becomes unstable. Since the electromagnetic fields created by the bunches are proportional to their charge, the onset of instabilities, and their amplitude are normally current-dependent.

Large-amplitude instabilities can cause beam loss, thus limiting the maximum stored current. If non-linear effects prevent beam loss by saturating the oscillation amplitude, instabilities degrade the beam quality and reduce the brightness or the luminosity of the accelerator.

### 1.2 Sources of coupled-bunch instabilities and cures

#### *1.2.1 Cavity high-order modes*

High-Q spurious resonances of the accelerating cavity structure are excited by the bunched beam, and act back on the beam itself. Each bunch affects the following bunches through the wake fields excited in the cavity. In this way the cavity high-order modes (HOMs) can couple with a beam oscillation mode having the same frequency and give rise to instability. Both transverse and longitudinal multi-bunch modes can be excited (Fig. 1).

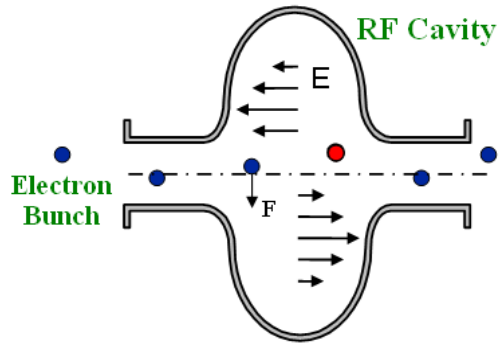

**Fig. 1:** RF cavity HOMs excited by a particle beam

Excitation of cavity HOMs can be avoided by a thorough design of the RF cavity or by the use of mode dampers made of antennas and resistive loads. Tuning of HOM frequencies through plungers or by changing the temperature of the cavity-cooling water is also currently used in many accelerators [1].

### 1.2.2 *Resistive wall impedance*

Resistive wall transverse instabilities are generated by the interaction of the beam with the vacuum chamber through the so-called 'skin effect' [2] (Fig. 2). They are particularly strong where low-gap chambers and in-vacuum insertion devices (undulators and wigglers) are used.

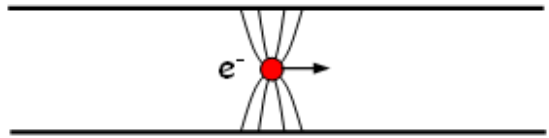

**Fig. 2:** Interaction of a bunch with the vacuum chamber

This effect can be reduced with an optimization of the vacuum chamber geometry and by using vacuum pipes made of low-resistivity materials.

### 1.2.3 *Interaction with vacuum chamber objects*

Coupled-bunch instabilities in both the transverse and longitudinal planes can be created by discontinuities of the vacuum chamber or small cavity-like structures located, for example, in beam position monitors (BPMs), vacuum pumps, bellows, etc. (Fig. 3).

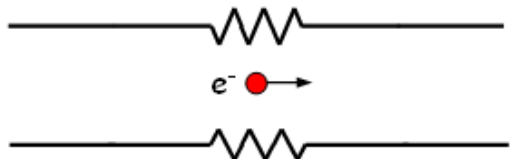

**Fig. 3:** Discontinuity of the vacuum chamber created by a bellows

The detrimental effect on coupled-bunch instabilities can be reduced with a proper design of the vacuum chamber and of the various installed objects [3].

## 1.2.4 Ion instabilities

Molecules of gas in the vacuum chamber can be ionized by collision with the electron beam. The generated positive ions remain trapped in the negative electric potential and produce electron–ion coherent oscillations [4].

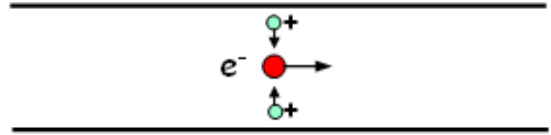

**Fig. 4:** Interaction of positive ions with electron bunches

A common practice currently used to counteract ion instabilities is so-called 'ion cleaning', obtained by operating the machine with a gap in the bunch train.

## 1.2.5 Generic countermeasures

Besides passive cures that reduce unwanted effects by acting directly on the sources of instability, it is possible to counteract the onset of coupled-bunch instabilities by increasing the Landau damping, which can be obtained by enhancing the betatron/synchrotron tune spread. This can be accomplished, for example, by acting on the chromaticity, by the use of higher harmonic RF cavities producing bunch lengthening [5] or by modulating the amplitude or phase of the accelerating radio frequency [6]. When all of the above-mentioned techniques are insufficient to reduce or eliminate coupled-bunch instabilities, active feedback mechanisms are required [7–12].

## 1.3 Equation of motion of particle bunches

### 1.3.1 Equation of motion of charged particles

The motion of a charged particle in a storage ring can be described with the harmonic oscillator analogy following the equation of motion

$$\ddot{x}(t) + 2D\dot{x}(t) + \omega^2 x(t) = 0 \tag{1}$$

where $x$ is the oscillation coordinate (transverse or longitudinal displacement), $D$ the natural damping, and $\omega$ the betatron/synchrotron tune frequency given by the $\nu\omega_0$ where $\nu$ is the betatron/synchrotron tune, and $\omega_0$ the revolution frequency. In the transverse planes, oscillations are periodic horizontal or vertical displacements from an ideal trajectory, while in the longitudinal plane they are periodic 'timing' differences with respect to a reference 'stable particle' running in the accelerator at constant speed. Longitudinal oscillations are often referred to as 'phase oscillations'.

If $\omega \gg D$, an approximated solution of the differential equation Eq. (1) is a damped sinusoidal oscillation

$$x(t) = k\, e^{-\frac{t}{\tau_D}} \sin(\omega t + \phi) \tag{2}$$

where $\tau_D = 1/D$ is the 'damping time constant' ($D$ is the damping rate).

Any excited oscillation (e.g. by quantum excitation) is damped by the natural damping (e.g. due to radiation of synchrotron light). The oscillation of individual particles is uncorrelated and manifests itself as an emittance growth.

*1.3.2 Coherent bunch oscillations*

Coupling with other bunches through the interaction with surrounding metallic structures adds a 'driving force' term $F(t)$ to the equation of motion

$$\ddot{x}(t)+2D\,\dot{x}(t)+\omega^2 x(t)=F(t) \quad (3)$$

Under given conditions, the oscillation of individual particles becomes correlated and the centroid of the bunch oscillates coherently with the other bunches giving rise to coherent bunch (coupled-bunch) oscillations. Each bunch oscillates according to the equation of motion:

$$\ddot{x}(t)+2(D-G)\,\dot{x}(t)+\omega^2 x(t)=0 \quad (4)$$

where $\tau_G = 1/G$ is the 'growth time constant' ($G$ is called the growth rate). Similarly to Eq. (2), if $\omega \gg (D - G)$, an approximated solution of Eq. (4) is

$$x(t)= k\,e^{-\frac{t}{\tau}} \sin(\omega t + \phi) \quad (5)$$

where $\dfrac{1}{\tau} = \dfrac{1}{\tau_D} - \dfrac{1}{\tau_G}$. If $D > G$ the oscillation amplitude decays; if $D < G$ it grows exponentially.

Since $G$ is proportional to the stored beam current, if the latter is lower than a given current threshold the beam remains stable; if higher a coupled-bunch instability is excited.

*1.3.3 Feedback action*

The feedback action adds a damping term $D_{fb}$ to the equation of motion

$$\ddot{x}(t)+2(D-G+D_{fb})\,\dot{x}(t)+\omega^2 x(t)=0 \quad (6)$$

In order to damp the oscillation $D_{fb}$ must be such that $D - G + D_{fb} > 0$.

A multi-bunch feedback system detects an instability by means of one or more BPMs and acts back on the beam by applying electromagnetic 'kicks' to the bunches.

Since the bunch oscillation is sinusoidal, the turn-by-turn position of a bunch measured at a given location is a sampled sinusoid. From Eq. (6), in order to introduce damping, the force applied by the feedback must be proportional to the derivative of the bunch oscillation. Consequently, the kick signal applied by the actuator to each bunch can be generated by shifting by $\pi/2$ the 'sampled' signal of the position of the same bunch when it passes through the 'kicker'. This is the basic principle of bunch-by-bunch feedback systems that will be treated in more detail in Section 2.

## 1.4 Multi-bunch modes

When a coupled-bunch instability is established, although each bunch oscillates at the tune frequency, there can be different modes of oscillation, called multi-bunch modes, depending on how each bunch oscillates with respect to the other bunches [9, 13].

Let us consider $M$ bunches equally spaced around the ring. Each multi-bunch mode is characterized by a bunch-to-bunch phase difference of

$$\Delta\Phi = m\,\frac{2\pi}{M} \quad (7)$$

where $m$ is the multi-bunch mode number ($m = 0, 1, ..., M - 1$). Each multi-bunch mode is associated to a characteristic set of frequencies

$$\omega = p\,M\,\omega_0 \pm (m+\nu)\,\omega_0 \quad (8)$$

where $p$ is an integer number ($-\infty < p < \infty$), $\omega_0$ is the revolution frequency, $\omega_{rf} = M\omega_0$ is the RF frequency (bunch repetition frequency), and $\nu$ is the tune. In turn, there are two sidebands at $\pm(m + \nu)\omega_0$ for each multiple of the RF frequency.

In the following sections we shall give some examples of transverse oscillation modes and examine the associated spectra.

### 1.4.1 Examples of transverse multi-bunch modes and their spectra

#### 1.4.1.1 One single stable bunch

In this example a single bunch of particles is stored in the ring. Figure 5 shows the linear trajectory of the bunch during its run along the ring (dotted line in Fig. 5(a)) and the time domain signal captured by a pickup (Fig. 5(b)). The location of the pickup is identified in the figure by triangles placed at longitudinal coordinate 0.

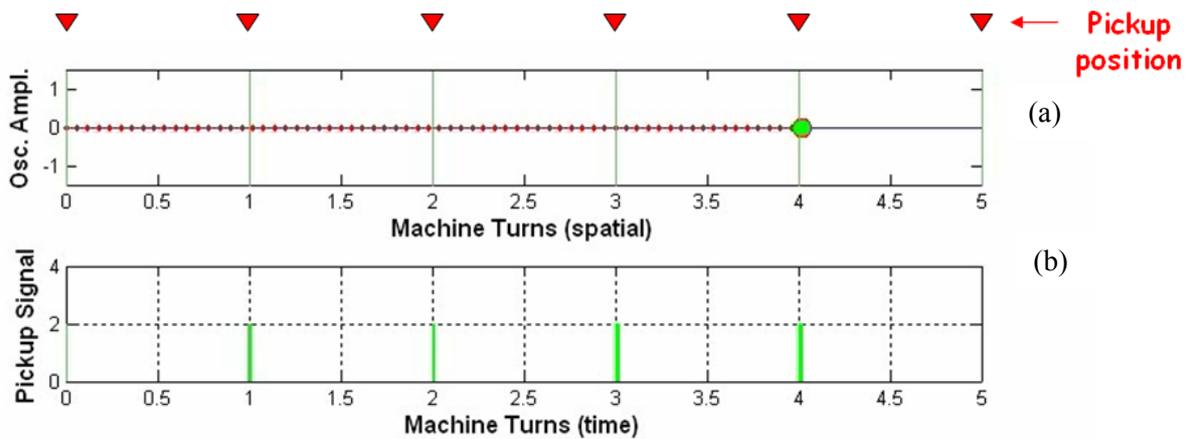

**Fig. 5:** (a) Motion of a single stable bunch; (b) time domain signal captured by a pickup

Every time the bunch passes through the pickup, an electrical pulse with constant amplitude is generated (vertical bars). If we think of it as a Dirac impulse, the spectrum of the pickup signal (Fig. 6) is a repetition of frequency lines at a multiple of the revolution frequency $p\omega_0$ for $-\infty < p < \infty$.

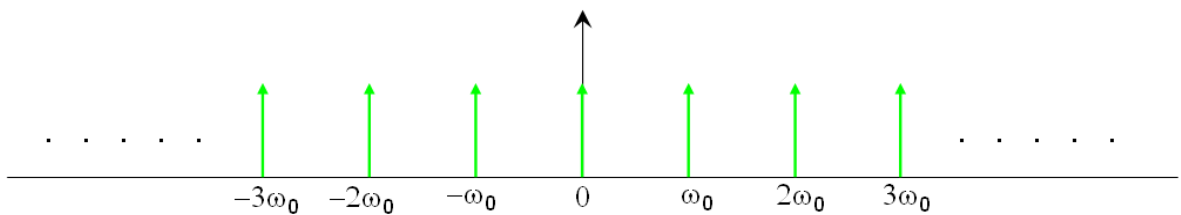

**Fig. 6:** Spectrum of the signal captured by the pickup with a single stable bunch

#### 1.4.1.2 One single unstable bunch

In this case a single bunch stored in the ring oscillates at the tune frequency $\nu\omega_0$. For simplicity, in this example we consider a tune value lower than unity, $\nu = 0.25$. With a single bunch $M = 1$, only mode number 0 exists. The bunch performs one complete oscillation in four machine turns and the pickup signal is a sequence of pulses modulated in amplitude with frequency $\nu\omega_0$ (Fig. 7).

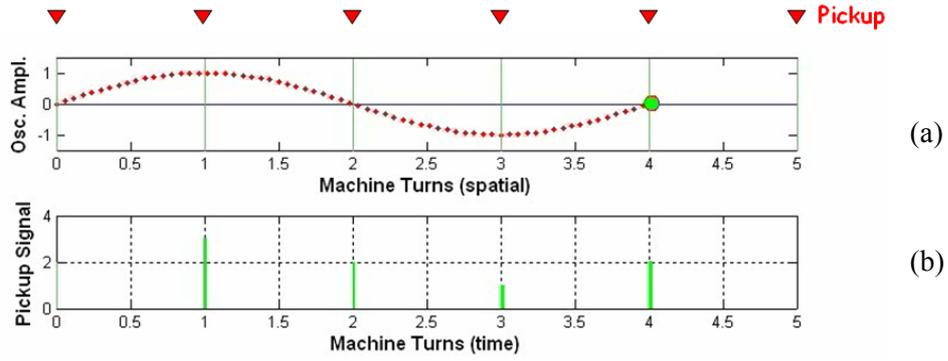

**Fig. 7:** (a) A single bunch oscillating at the tune frequency; (b) corresponding pickup signal

In the spectrum two sidebands at $\pm \nu\omega_0$ appear at each of the revolution harmonics (Fig. 8).

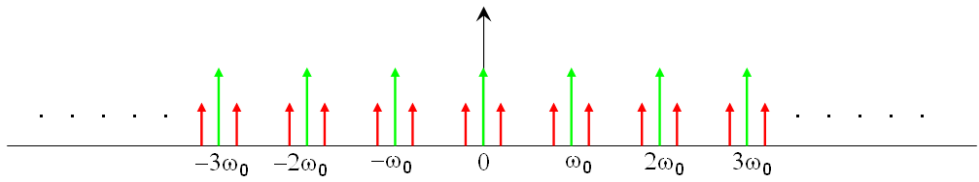

**Fig. 8:** Spectrum of the pickup signal with a single unstable bunch

*1.4.1.3   Ten stable bunches*

Let us consider now ten identical, equally spaced, stable bunches ($M = 10$). The pickup signal is a sequence of pulses at the bunch repetition frequency (RF frequency) $\omega_{rf} = 10\,\omega_0$.

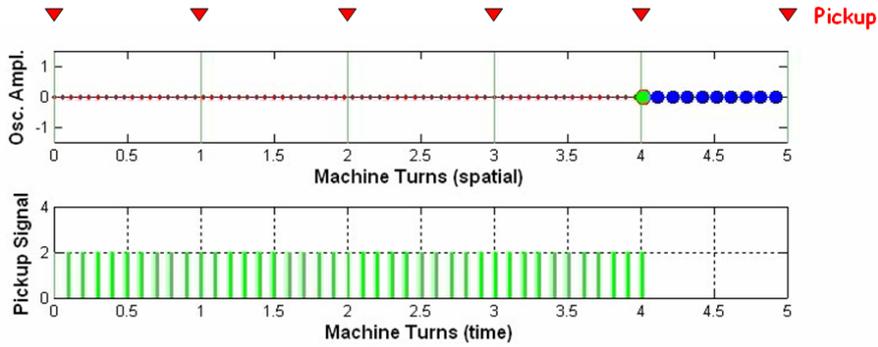

**Fig. 9:** Ten equally spaced stable bunches

The spectrum is a repetition of frequency lines at multiples of the bunch repetition frequency.

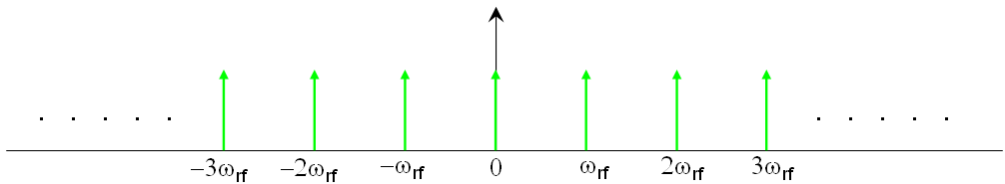

**Fig. 10:** Spectrum of the pickup signal with ten stable bunches

*1.4.1.4   Ten unstable bunches, mode #0*

In the case where the bunches oscillate at the tune frequency $\nu\omega_0$ with $\nu = 0.25$, since $M = 10$ there are 10 possible modes of oscillation characterized by a phase difference between adjacent bunches of $\Delta\Phi = m\,\dfrac{2\pi}{10}$ with $m = 0, 1, ..., 9$.

If *m* = 0 (mode number 0) then ΔΦ = 0, namely all the bunches oscillate with the same phase (Fig. 11). The pickup signal is a sequence of pulses at the bunch repetition frequency modulated in amplitude with frequency $\nu\omega_0$.

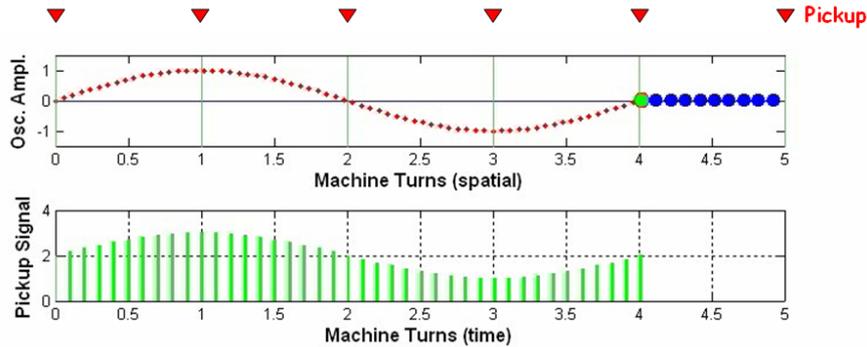

**Fig. 11:** Ten identical bunches oscillating in phase at the tune frequency. The dotted line is the trajectory of one of them.

The spectrum is a repetition of frequency lines at multiples of the bunch repetition frequency with sidebands at $\pm\nu\omega_0$: $\omega = p\omega_{rf} \pm \nu\omega_0$ with $-\infty < p < \infty$ (Fig. 12).

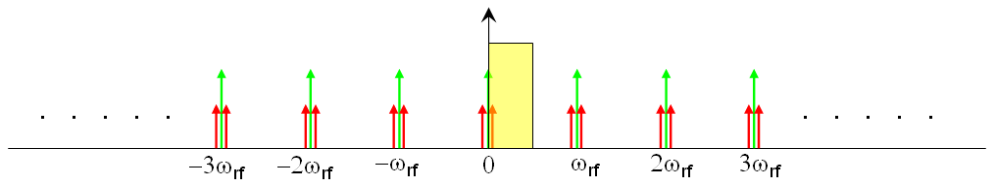

**Fig. 12:** Spectrum of ten unstable bunches: mode #0

Since the spectrum is periodic and the mode appears twice (lower and upper sideband) in a $\omega_{rf}$ frequency span centred on every harmonic of $\omega_{rf}$, we can limit the spectrum analysis to a $0-\omega_{rf}/2$ frequency range. Figure 13 is an enlargement of the frequency portion denoted by the rectangular box in Fig. 12.

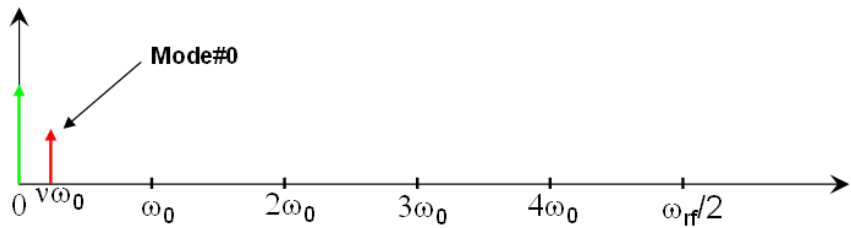

**Fig. 13:** Spectrum of mode #0 restricted to a $\omega_{rf}/2$ wide frequency range

*1.4.1.5  Ten unstable bunches, mode #1*

In mode #1, the phase difference of the bunches oscillating at the betatron frequency is $\Delta\Phi = 2\pi/10$ and the spectrum is characterized by frequency lines at $\omega = p\omega_{rf} \pm (\nu + 1)\omega_0$, with $-\infty < p < \infty$ (Figs. 14 and 15).

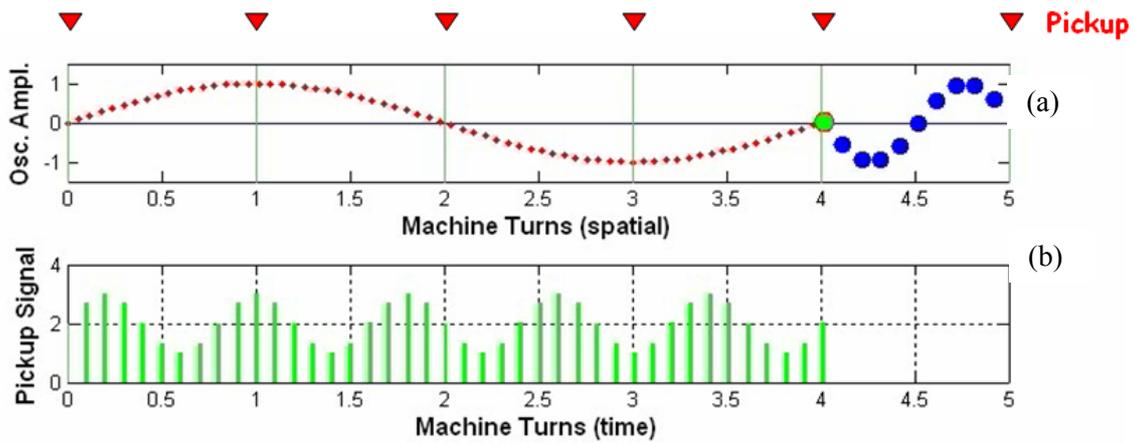

**Fig. 14:** Ten unstable bunches: mode #1

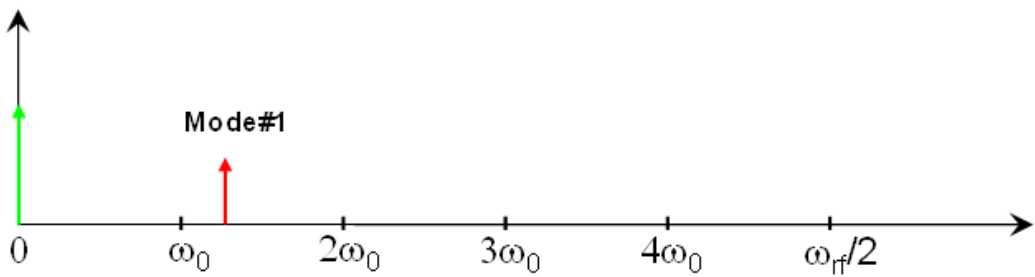

**Fig. 15:** Spectrum of mode #1

*1.4.1.6  Ten unstable bunches, modes #2 to #9*

Figures 16–23 show the motion, time domain signal, and spectrum of multi-bunch modes #2 to #9.

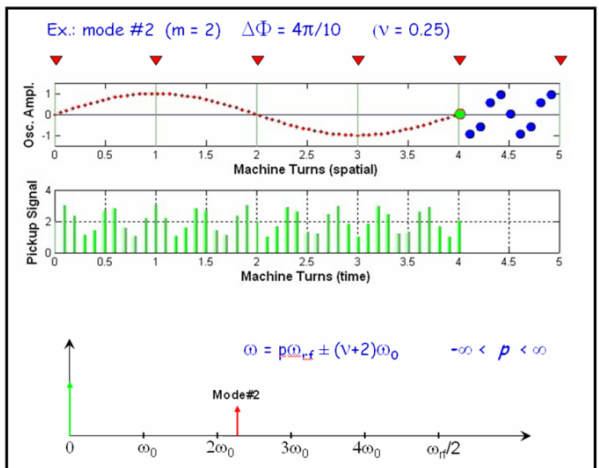      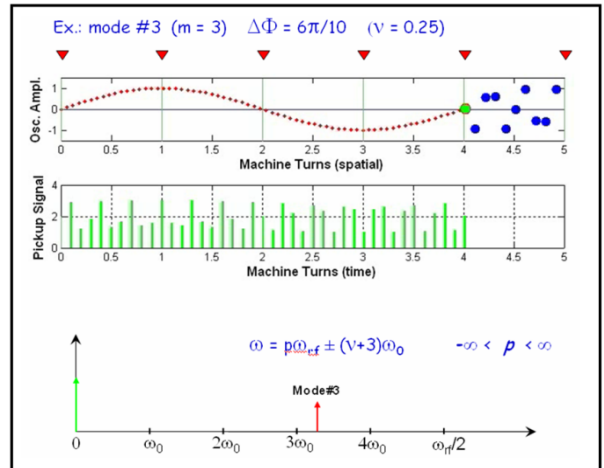

**Fig. 16:** Mode #2                                    **Fig. 17:** Mode #3

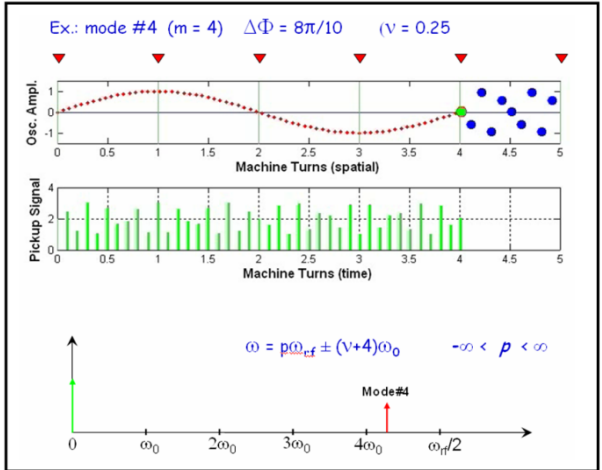

**Fig. 18:** Mode #4

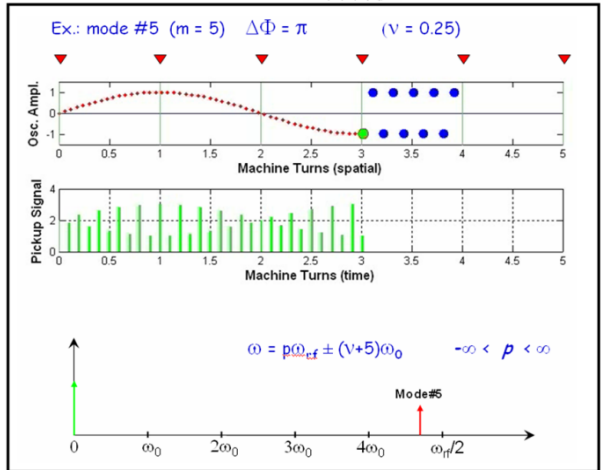

**Fig. 19:** Mode #5

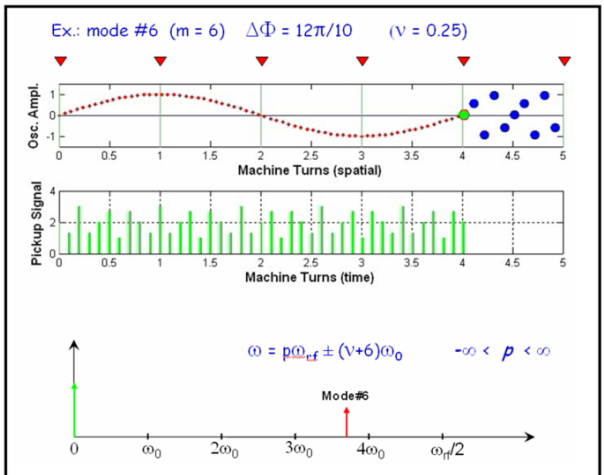

**Fig. 20:** Mode #6

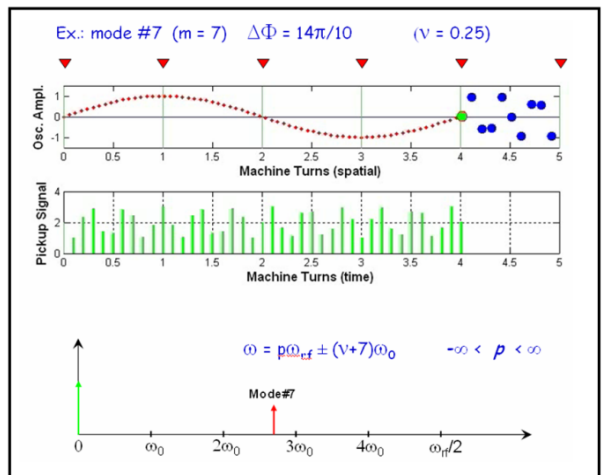

**Fig. 21:** Mode #7

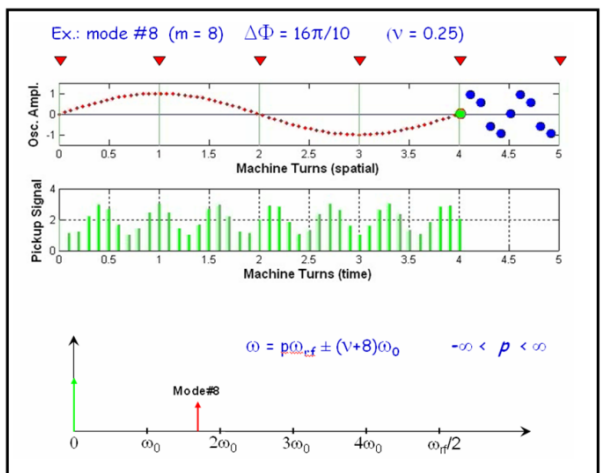

**Fig. 22:** Mode #8

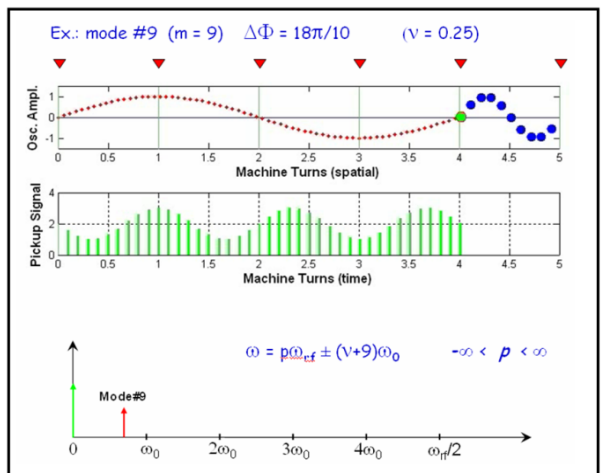

**Fig. 23:** Mode #9

### 1.4.2    Uneven filling

If the bunches do not have the same charge, i.e. the buckets are not equally filled (uneven filling), the spectrum also has frequency components at the revolution harmonics (multiples of $\omega_0$) (Fig. 24). The amplitude of the revolution harmonics depends on the filling pattern of one machine turn.

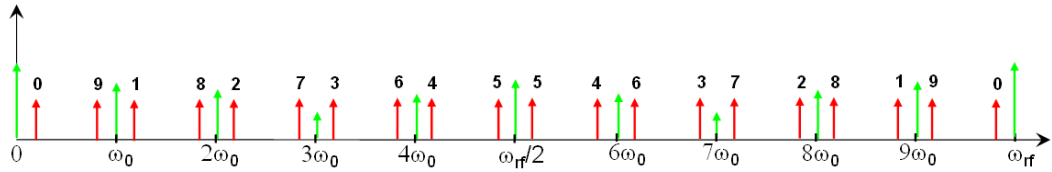

**Fig. 24:** Spectrum of a beam with uneven filled buckets

### 1.4.3 Longitudinal multi-bunch modes

Compared with transverse oscillations where multi-bunch instabilities produce amplitude modulation of the stable beam pickup signal, longitudinal multi-bunch modes produce phase modulation. Components at $\pm \nu \omega_0$, $\pm 2\nu \omega_0$, $\pm 3\nu \omega_0 \ldots$ appear aside the revolution harmonics. Their amplitude depends on the depth of the phase modulation, namely the amplitude of the instability, according to Bessel series expansion (Fig. 25).

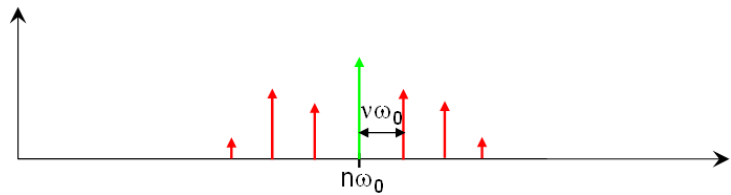

**Fig. 25:** Spectrum of a longitudinally unstable beam with components near to a revolution harmonic at $\pm \nu \omega_0$, $\pm 2\nu \omega_0$, $\pm 3\nu \omega_0 \ldots$.

### 1.4.4 Coupled-bunch instability

Starting from the spectral representation of all potential multi-bunch modes, one of them can become unstable if one of its sidebands overlaps, for example, with the frequency response of a HOM. The HOM couples with the sideband giving rise to a coupled-bunch instability, with a consequent increase of the mode amplitude (Fig. 26).

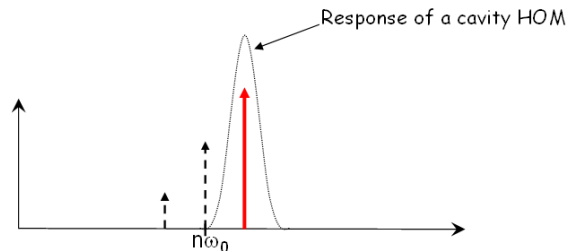

**Fig. 26:** Coupling of a cavity HOM with the sideband corresponding to a given multi-bunch mode

More generally, in order to characterize the interaction of the beam with the machine structures, the 'impedance model' is often used. The impedance is the frequency spectrum of the forces generated by wake fields acting back on the beam. It summarizes the electromagnetic effects of the machine components (vacuum chamber, cavities, bellows, BPMs, etc.) on the beam. The shape and amplitude of the machine impedance give us information about which multi-bunch modes could be excited and the strength of the excitation (growth time constant).

### 1.4.5 Effects of coupled-bunch instabilities

Besides the risk of beam loss or the limitation of the maximum stored current, the onset of instabilities produces an increase of the transverse beam size (also in the case of longitudinal instabilities because of dispersion) and thus of the effective emittance. This is detrimental for the beam quality since it lowers the brightness of the produced photon beams in synchrotron light sources and the luminosity in circular colliders. Besides these negative effects, however, the presence of multi-bunch instabilities is often associated with an increase of the beam lifetime. Saturation effects due to Landau damping, in fact, produce dilution of particles inside the bunches with a consequent decrease of Touschek scattering.

### 1.4.6  *Real examples of multi-bunch modes and their measurement*

Let us consider two real examples of transverse and longitudinal instabilities. We refer to the Elettra synchrotron storage ring. The RF frequency is 499.654 MHz; having a harmonic number of 432, the revolution frequency is 1.157 MHz. The horizontal tune $\nu_{hor}$ is 12.30 (fractional tune frequency = 345 kHz), the vertical tune $\nu_{ver}$ is 8.17 (fractional tune frequency = 200 kHz), and the longitudinal tune $\nu_{long}$ is 0.0076 (tune frequency = 8.8 kHz). Keeping in mind Eq. (8) ( $\omega = p\,M\,\omega_0 \pm (m+i)\,\omega_0$ ) let us analyse two beam spectra measured with a spectrum analyser connected to a strip line pickup.

#### 1.4.6.1  *Example 1*

In this example we analyse a spectral line at 512.185 MHz (Fig. 27). It is a lower sideband of the second harmonic of the RF frequency, 200 kHz apart from the 443rd revolution harmonic at 512.355 MHz, denoting that it is a vertical instability. From Eq. (8) it can be easily calculated that it corresponds to the vertical multi-bunch mode #413.

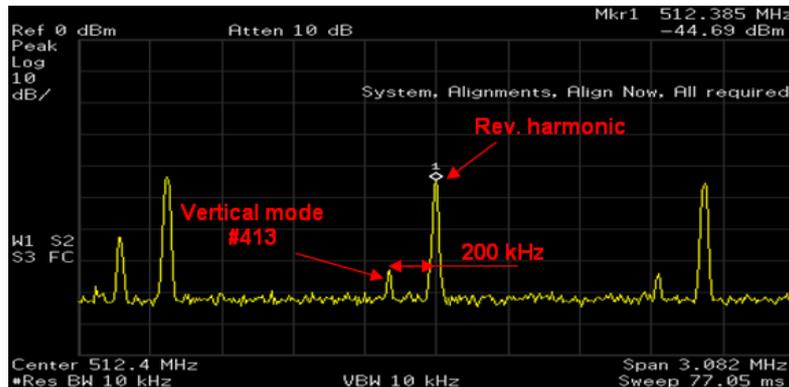

**Fig. 27:** Example of beam spectrum in the presence of vertical instabilities measured by a spectrum analyser

#### 1.4.6.2  *Example 2*

In this case we analyse a spectral line at 604.914 MHz (Fig. 28). It is an upper sideband of the RF frequency, 8.8 kHz apart from the 523rd revolution harmonic, denoting a longitudinal instability. From Eq. (8) we can calculate that it corresponds to the longitudinal multi-bunch mode #91.

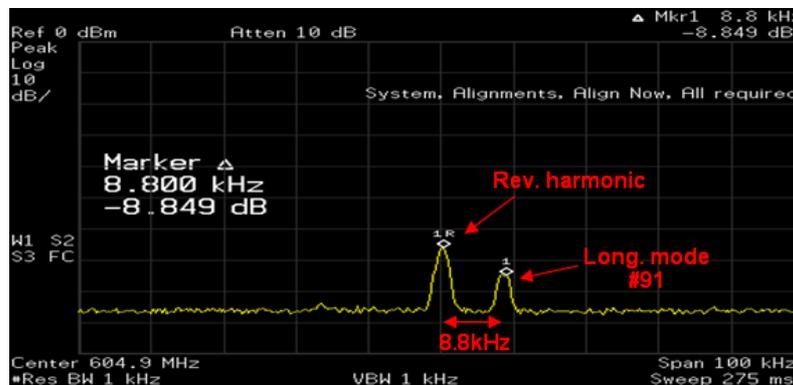

**Fig. 28:** Example of beam spectrum in the presence of a longitudinal instability measured by a spectrum analyser

## 2 Multi-bunch feedback systems

### 2.1 Introduction

As already seen in Section 1.3.3, a multi-bunch feedback system detects multi-bunch instabilities using one or more BPMs and acts back on the beam through an electromagnetic actuator called a 'kicker'. A simplified block diagram of a feedback system is depicted in Fig. 29.

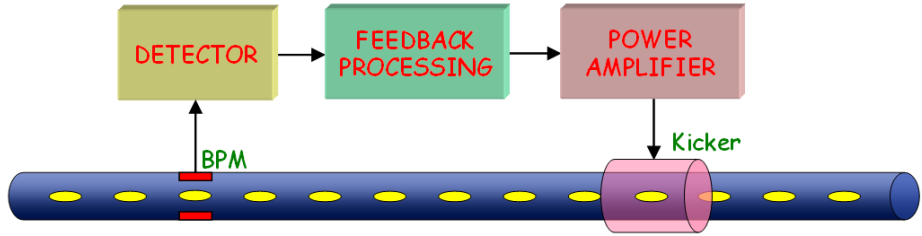

**Fig. 29:** Block diagram of a multi-bunch feedback system

BPMs and detectors measure the oscillation of the bunches and the processing unit generates the correction signal, which is amplified by the power amplifier and sent to the kicker.

### 2.2 Types of feedback systems

Depending on how a feedback system detects and controls instabilities, there are two main feedback categories: 'mode-by-mode' and 'bunch-by-bunch' feedback. The former are often referred to as 'frequency domain' feedback, the latter as 'time domain' feedback [8]. Moreover, the feedback implementation can be analogue or digital, depending on whether the signal is handled in the analogue domain or sampled and processed digitally. Despite the fact that most of the latest feedback systems are digital bunch-by-bunch systems, analogue or mode-by-mode implementations are still in use in a number of accelerators.

In the following sections the basic principles of the above-mentioned types of feedback will be illustrated; they are part of the history of multi-bunch feedback systems over a period of at least thirty years.

#### 2.2.1 *Mode-by-mode feedback*

Mode-by-mode (frequency domain) feedback acts separately on each of the controlled unstable modes [14]. A simplified block diagram is illustrated in Fig. 30.

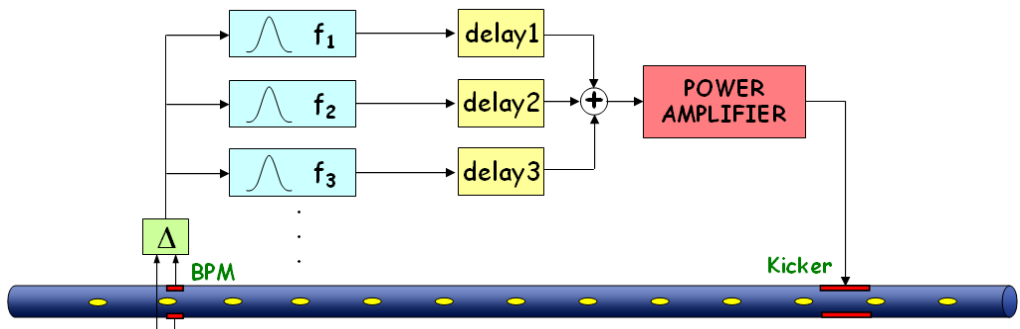

**Fig. 30:** Block diagram of a mode-by-mode feedback system

An appropriate electronic system generates the position error signal by combining the signals of the BPM buttons. A number of processing chains, each dedicated to one of the controlled modes, work in parallel. These are composed of a narrow band-pass filter that selects the proper frequency and an adjustable delay line to phase-shift the signal in order to produce a negative feedback. The channels are then combined, amplified, and sent to the kicker.

### 2.2.2 Bunch-by-bunch feedback

A bunch-by-bunch (time domain) feedback individually steers each bunch by applying 'small' electromagnetic kicks every time the bunch passes through the kicker. The correction signal for a given bunch is generated based on the motion of that bunch. The result is a damped oscillation lasting several turns. Bunch-by-bunch feedback can be realized as shown in Fig. 31.

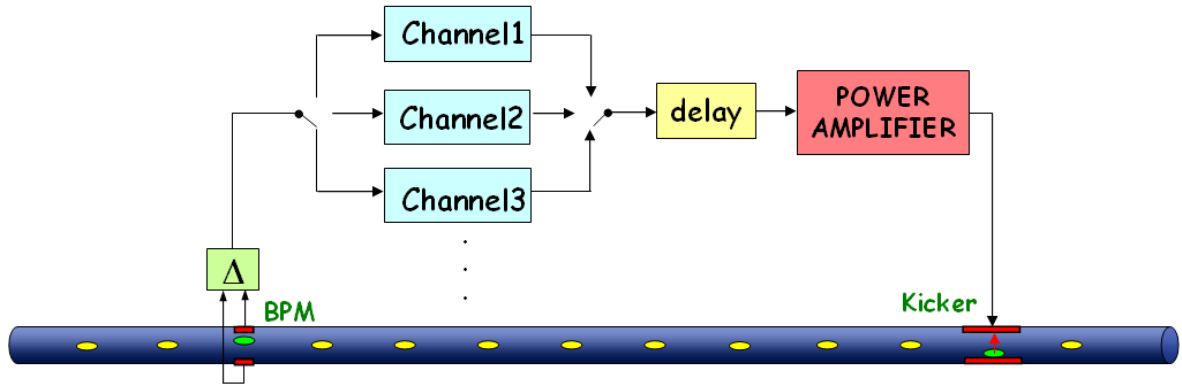

**Fig. 31:** Block diagram of a bunch-by-bunch feedback system based on a time division scheme

There are as many processing channels as the number of bunches. A demultiplexer provides each channel with the samples for the corresponding bunch, while a multiplexer recombines the samples at the end of the channels. Following the principle explained in Section 1.3.3, the main task of each channel is to properly phase-shift the sampled position signal. A delay line assures that each bunch is kicked with its own correction samples.

Every bunch is measured and corrected at every machine turn but, owing to the delay of the feedback chain, the correction kick corresponding to a given measurement is applied to the bunch one or more turns later.

An important principle behind bunch-by-bunch feedback is that it is possible to stabilize all coupled-bunch modes by damping the oscillation of each bunch.

### 2.2.3 Analogue bunch-by-bunch feedback

In the following sections we shall see some real analogue implementations of transverse bunch-by-bunch feedback, starting from a basic very simple architecture to a system that has been deployed in many accelerators.

#### 2.2.3.1 Feedback using one BPM

In bunch-by-bunch feedback the correction signal applied to a given bunch must be proportional to the derivative of the bunch oscillation at the kicker. The correction signal is therefore a sampled sinusoid shifted by π/2 with respect to the oscillation of the bunch when it passes through the kicker. Starting from this principle, a very simple way to obtain such a phase shift is to have the BPM and kicker placed at points in the ring with the appropriate betatron phase difference and to use the detected signal to directly feed the kicker through an amplifier. The system is shown in Fig. 32.

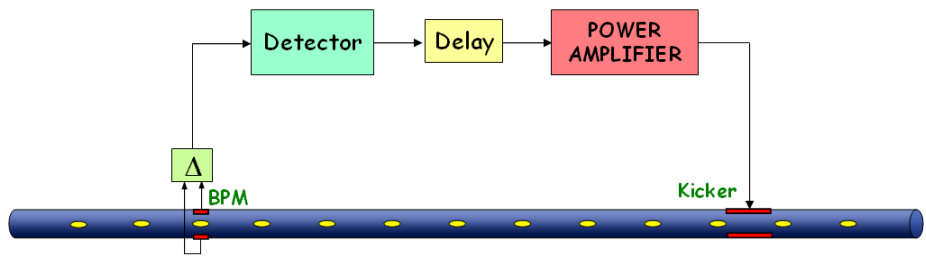

**Fig. 32:** Block diagram of a simple analogue bunch-by-bunch transverse feedback system

The detector down-converts the high frequency (typically a multiple of the bunch frequency $f_{rf}$) BPM signal into baseband (range $0$–$f_{rf}/2$). The delay line assures that the signal for a given bunch passing through the feedback chain arrives at the kicker when, after one machine turn, the same bunch again passes through it.

#### 2.2.3.2 Feedback using two BPMs

By using two BPMs, provided that they are separated by about $\pi/2$ in betatron phase, it is possible to place them in any ring position with respect to the kicker (Fig. 33).

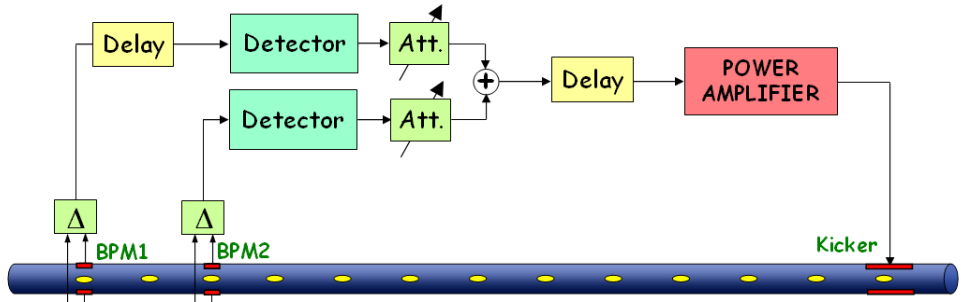

**Fig. 33:** Analogue bunch-by-bunch transverse feedback using two BPMs

As shown in Fig. 34, if we opportunely combine the two detected signals using variable attenuators, it is possible to vary the phase of the resulting bunch correction signal. The attenuators have to be adjusted so that the bunch oscillation and the corresponding kick signal are in quadrature.

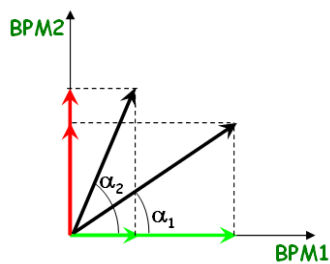

**Fig. 34:** If the two BPM bunch signals are in quadrature, the phase of the sum signal can be varied by adjusting the variable attenuators.

#### 2.2.3.3 Suppression of revolution harmonics

The revolution harmonics (frequencies at multiples of $\omega_0$) are useless components that can be removed from the correction signal in order not to saturate the RF amplifier. A special module implementing notch filters at the harmonics of $\omega_0$ can be placed before the amplifier to eliminate these components. This operation is also called 'stable beam' rejection. We shall see below a number of possible implementations of this module that are also used in digital feedback systems.

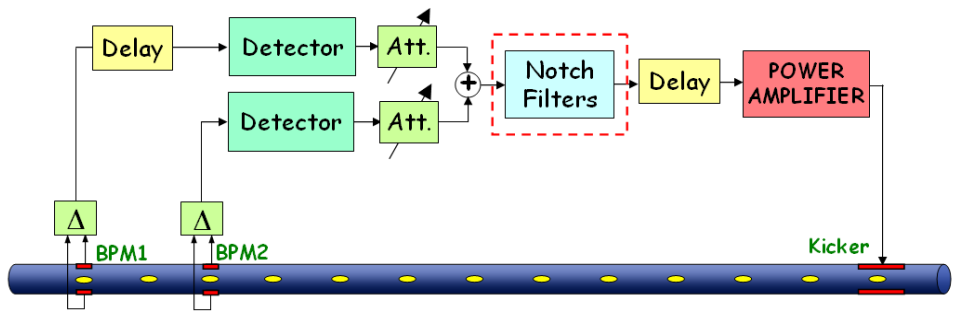

**Fig. 35:** Analogue bunch-by-bunch transverse feedback with suppression of revolution harmonics

Architectures similar to the above examples have been used to build the ALS [15], Bessy II [16], PLS [17] and ANKA [18] transverse multi-bunch feedback systems.

### 2.2.4 Digital bunch-by-bunch feedback

Commercially available fast digital electronics (analogue-to-digital converters (ADCs), digital-to-analogue converters (DACs), digital signal processors (DSPs), field programmable gate arrays (FPGAs), etc.) allow one to build digital feedback systems where the signal of the bunch positions is digitized, digitally processed to calculate the corrections, and re-converted to an analogue signal. A block diagram of a generic digital feedback is illustrated in Fig. 36. Thanks to the capabilities offered by digital signal processing, digital feedback usually makes use of only one BPM and a kicker mounted in any ring position.

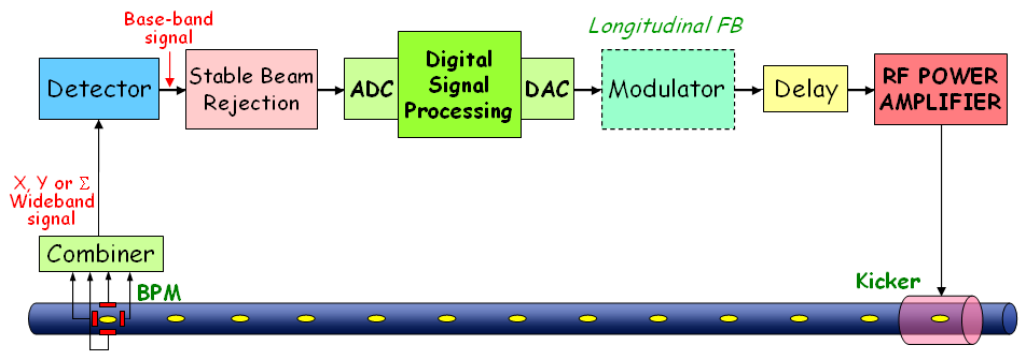

**Fig. 36:** Block diagram of a digital bunch-by-bunch feedback system

Starting from the BPM button signals, the combiner generates the wideband horizontal (X), vertical (Y) or sum ($\Sigma$) signal, which is then demodulated to baseband by the detector (also called 'RF front-end'). Some feedback implementations feature a stable beam rejection module that removes useless stable beam components from the signal, which is eventually digitized, processed, and re-converted to analogue by the digital processor. While in transverse feedback systems the amplifier and kicker operate in baseband, longitudinal feedback requires a modulator that translates the correction signal to the kicker operation frequency. The delay line adjusts the timing of the signal to match the bunch arrival time.

### 2.2.5 Digital vs. analogue feedback

The following list summarizes the main advantages of digital feedback systems:

i) reproducibility: when the signal is digitized it is not subject to temperature/environment changes or ageing;

ii) programmability: the implementation of processing functionalities is usually made using DSPs or FPGAs, which are programmable via software/firmware;

iii) performance: digital controllers feature superior processing capabilities with the possibility to implement sophisticated control algorithms not feasible in analogue;

iv) additional features: the possibility to combine basic control algorithms and additional useful features like signal conditioning, saturation control, down-sampling, etc.;

v) implementation of diagnostic tools, used for both feedback commissioning and machine physics studies (see Section 6);

vi) easier and more efficient integration of the feedback in the accelerator control system, important for feedback set-up and tuning, fast data acquisition, easy and automated operations, etc.

Among the disadvantages is the higher delay of the feedback chain (due to ADC, digital processing, and DAC) with respect to analogue feedback, although with the use of FPGAs this delay is reduced to acceptable values.

## 3 Components of feedback systems

In the following sections a more detailed description of components usually employed in (digital) feedback systems will be given, with practical issues, examples, and real implementations.

### 3.1 BPM and combiner

The four signals from a standard four-button BPM can be opportunely combined to obtain the wideband X, Y and Σ signals used by the horizontal, vertical, and longitudinal feedback systems, respectively. Figure 37 shows an example of how this can be made.

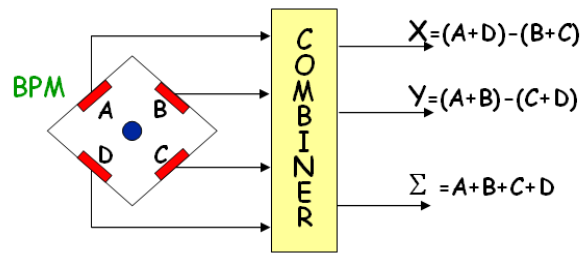

**Fig. 37:** BPM and combiner are used to provide X, Y, and Σ wideband signals

Any $f_{rf}/2$ portion of the beam spectrum contains the information for all potential multi-bunch modes and can be used to detect instabilities and measure their amplitude. Usually the BPM and combiner work in a frequency range around a multiple of $f_{rf}$ (Fig. 38), where the amplitude of the overall frequency response of the BPM and cables is at a maximum. Moreover, a higher $f_{rf}$ harmonic is preferred for the longitudinal feedback because of the better sensitivity of the phase detection system.

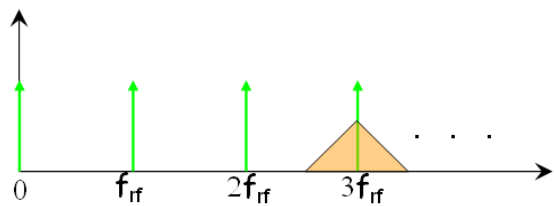

**Fig. 38:** Example of wideband signal centred on the third harmonic of the RF frequency

## 3.2 Detector (RF front-end)

### 3.2.1 Transverse case

The detector translates the wideband signal to baseband (0–$f_{rf}/2$ range): the operation is an amplitude demodulation (Fig. 39).

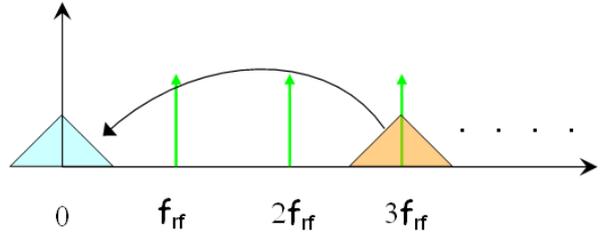

**Fig. 39:** The amplitude demodulation translates the wideband signal to baseband

A possible implementation of amplitude demodulation, widely used in telecommunication technology, is the 'heterodyne' technique illustrated in Fig. 40. The wideband signal is first band-pass filtered to select the desired frequency range and then mixed with the 'local oscillator' signal, which must be synchronous with the former. The local oscillator signal can be derived from the RF by multiplying its frequency by an integer number corresponding to the chosen harmonic of $f_{rf}$. The resulting signal is eventually low-pass filtered in the 0–$f_{rf}/2$ frequency range.

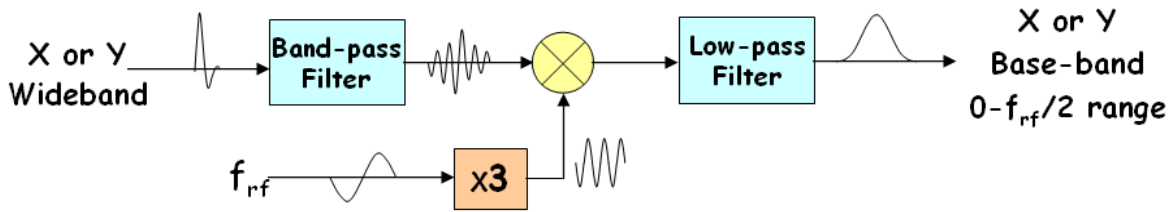

**Fig. 40:** Heterodyne amplitude demodulation using the third harmonic of the RF frequency

### 3.2.2 Longitudinal case

The wideband sum signal (Σ) contains only information about the phase (longitudinal position) of the bunches, since the sum of the four button signals has almost constant amplitude.

The baseband phase error signal (0–$f_{rf}/2$ range) can be generated by phase demodulation, which can be carried out with the same heterodyne technique used for amplitude demodulation but using a 'local oscillator' in quadrature with the wideband signal, namely shifted in phase by $\pi/2$ (Fig. 41).

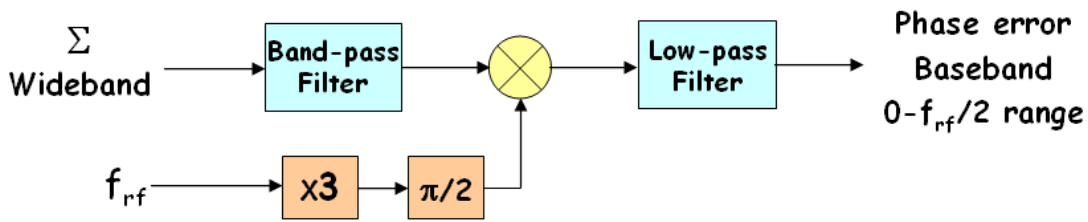

**Fig. 41:** Phase demodulation using the heterodyne technique

### 3.2.3 Amplitude and phase demodulation

The following are simple mathematic demonstrations that can help in understanding heterodyne amplitude and phase demodulations.

### 3.2.3.1 Amplitude demodulation

Let us consider a sinusoidal carrier $\sin(3\omega_{rf} t)$ modulated in amplitude by a signal $A(t)$. If we multiply it by the local oscillator signal $\sin(3\omega_{rf} t)$ (mixing operation) we obtain two frequency components. The component at the higher frequency can be filtered out by a low-pass filter leaving a signal proportional to the modulating signal $A(t)$

$$A(t) \sin(3\omega_{rf} t) \sin(3\omega_{rf} t) \propto A(t) (\cos(0) - \cos(6\omega_{rf} t)) \approx A(t) . \qquad (9)$$

### 3.2.3.2 Phase demodulation

In this case the sinusoidal carrier is phase modulated by a signal $\varphi(t)$. If the latter is small, by multiplying the modulated signal by a quadrature-phase local oscillator signal $\cos(3\omega_{rf} t)$, after low-pass filtering we obtain the modulating signal

$$\sin(3\omega_{rf} t + \varphi(t)) \cos(3\omega_{rf} t) \propto \sin(6\omega_{rf} t + \varphi(t)) + \sin(\varphi(t)) \approx \varphi(t) . \qquad (10)$$

### 3.2.4 Time domain considerations

The baseband signal can be seen as a sequence of 'pulses', each with amplitude proportional to the position error (X, Y or Φ) and to the charge of the corresponding bunch. By sampling this signal with an A/D converter synchronous with the bunch frequency, one can measure X, Y or Φ (Fig. 42).

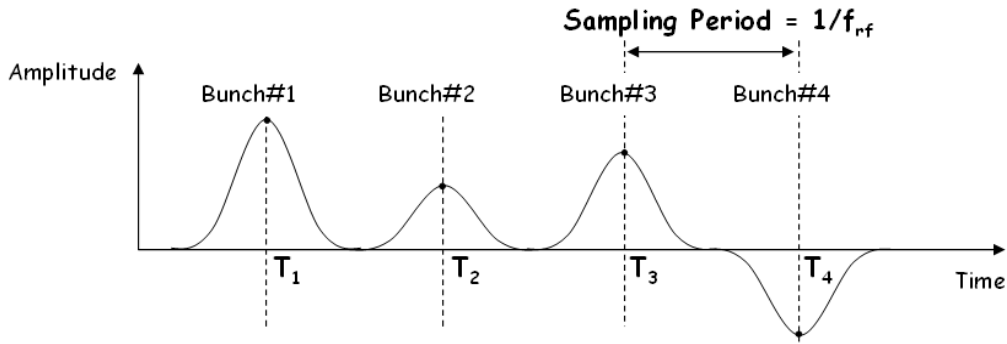

**Fig. 42:** Detector output signal

Note that the multi-bunch-mode number $M/2$, which is the one with the highest frequency (close to $f_{rf}/2$) in baseband, is characterized by a sequence of pulses with almost the same amplitude but alternating signs.

The design of the detector band-pass and low-pass filters affects the shape of the pulses that should have maximum flatness on the top (to reduce clock jitter noise) and minimum overlap with adjacent pulses (to reduce cross-talk between bunches) (Fig. 43). To achieve this, Bessel filters with linear phase response or a special filter called a 'comb generator' [19] can be used.

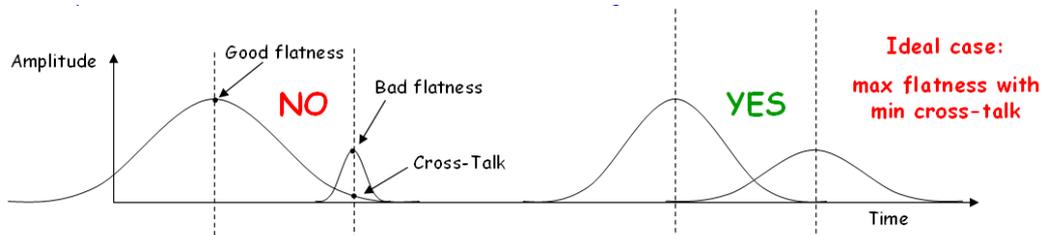

**Fig. 43:** Examples of badly and well-filtered detector signals

### 3.3 Rejection of stable beam signal

The amplitude of the pulses in the baseband signal can have a constant offset called the 'stable beam signal'. In the transverse plane this offset can be due to an off-centre beam in the BPM or to unbalanced BPM electrodes or cables. In the longitudinal plane it can be generated by a mismatch of the local oscillator phase with respect to the bunch phases, which is difficult to avoid when the bunches have different synchronous phases due to beam loading.

In the frequency domain, the presence of a stable beam signal carries non-zero revolution harmonics.

The stable beam signal is useless for feedback since it does not contain information about multi-bunch modes. It is therefore preferable to reduce it in order to avoid saturation of the ADC, DAC or amplifier. A number of techniques have been adopted to reduce this signal; we shall mention three examples.

### 3.3.1 Balancing of BPM buttons

In the case of transverse feedback, variable attenuators placed on the BPM electrodes can be used to equalize the amplitude of the signals (Fig. 44). In this way closed-orbit offsets or unbalanced signals can be compensated for.

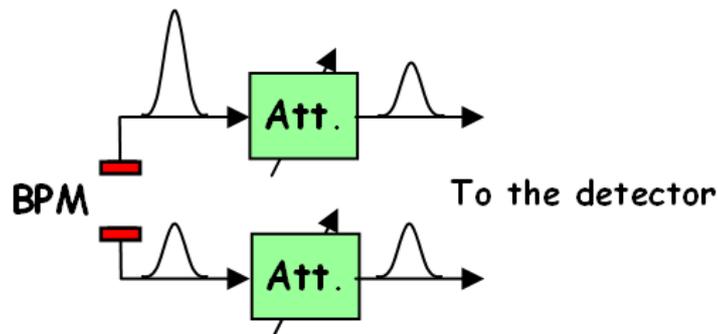

**Fig. 44:** Example of BPM signal balancing

### 3.3.2 Comb filter

An analogue filter performing stable beam rejection through the reduction of the revolution harmonics can be simply implemented by means of delay lines and combiners, as shown in Fig. 45(a). Its frequency response (Fig. 45(b), shows a series of notches at multiples of $\omega_0$ able to suppress all the revolution harmonics (DC included) from the detector output signal [18].

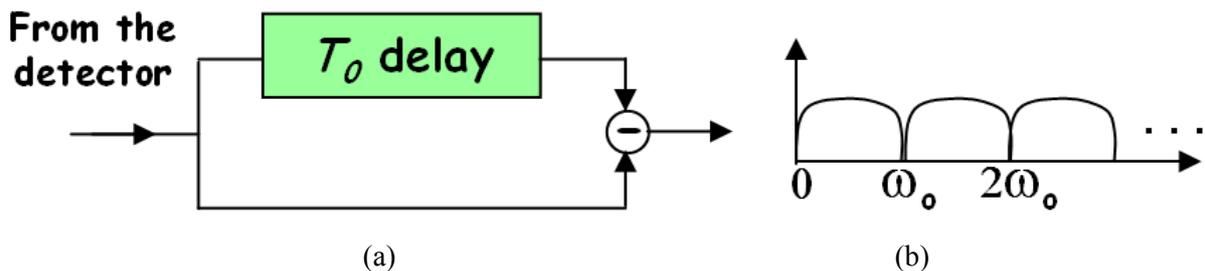

(a)                  (b)

**Fig. 45:** Stable beam rejection using a comb filter. (a) Block diagram of the filter; (b) frequency response

### 3.3.3 Digital DC rejection

The stable beam signal can be eliminated by removing the DC component from the turn-by-turn signal of every bunch. This can be done digitally (Fig. 46).

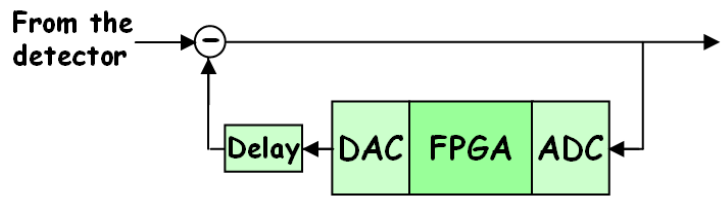

**Fig. 46:** Block diagram of a digital DC rejection system

The detector output signal is sampled by an ADC clocked at $f_{rf}$. The turn-by-turn digital signal of each bunch is integrated by a digital processing unit (e.g. a FPGA), recombined with the other bunch signals, converted to analogue, and subtracted from the original signal. In this way the constant component of each bunch is eliminated [20].

## 3.4 Digital processor

The A/D converter samples and digitizes the detector signal at the bunch repetition frequency; each sample corresponds to the position error (X, Y or Φ) of a given bunch. Precise synchronization of the sampling clock with the bunch signal must be provided.

The digital samples are then de-multiplexed into $M$ channels, $M$ being the number of bunches in the ring. In each channel the turn-by-turn samples of a given bunch are processed by a dedicated digital filter to calculate the correction samples. Basic processing consists of DC component suppression (if not completely accomplished by the external stable beam rejection) and phase shift of the signal at the betatron/synchrotron frequency (see Section 1.3.3).

The correction sample streams are then recombined and converted to analogue by the D/A converter (Fig. 47).

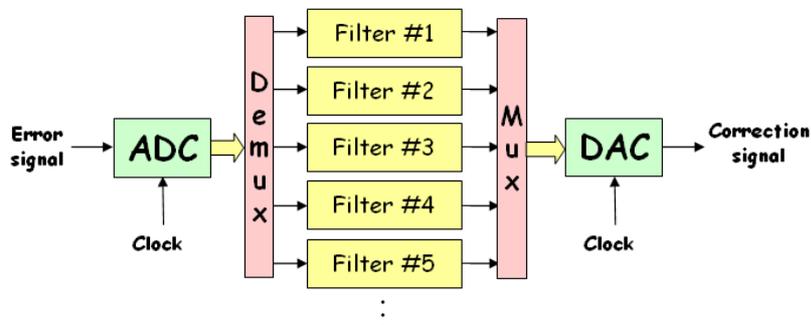

**Fig. 47:** Block diagram of bunch-by-bunch feedback by digital processor

### *3.4.1 Digital processor implementation*

#### 3.4.1.1 ADC

Existing multi-bunch feedback systems usually employ 8-bit ADCs clocked at up to 500 Msample/s; some implementations use a number of ADCs with higher resolution (e.g. 14 bits) and lower clock rate working in parallel. ADCs with enhanced resolution have the advantage of a lower quantization noise (crucial for low-emittance machines) and a higher dynamic range, which allows, in some cases, the avoidance of external stable beam rejection.

#### 3.4.1.2 DAC

The DACs usually employed convert samples at up to 500 Msample/s and 14-bit resolution.

### 3.4.1.3 *Digital processing*

Feedback processing can be performed by discrete digital electronics (an obsolete technology), DSPs or FPGAs. Figure 48 summarizes the advantages and disadvantages of the last two options. Although there are still some digital multi-bunch feedback systems employing DSPs, the trend is towards solutions using FPGAs.

|  | Pros | Cons |
|---|---|---|
| DSP | ➢ Easy programming<br>➢ Flexible | ➢ Difficult HW integration<br>➢ Latency<br>➢ Sequential program execution<br>➢ A number of DSPs are necessary |
| FPGA | ➢ Fast (only one FPGA is necessary)<br>➢ Parallel processing<br>➢ Low latency | ➢ Trickier programming<br>➢ Less flexible |

**Fig. 48:** Summary of advantages and disadvantaged of DSPs and FPGAs

### 3.4.2 *Examples of digital processors*

The following are examples of digital processor implementation employed in a number of accelerators across the world. We would like to point out that this list is probably incomplete and not up-to-date.

#### 3.4.2.1 *PETRA transverse and longitudinal feedback*

This is one of the first digital implementations of bunch-by-bunch feedback [21]. The digital processor is composed of an ADC, digital processing electronics made of discrete components (adders, multipliers, shift registers, etc.) implementing a FIR filter and a DAC.

#### 3.4.2.2 *ALS/PEP-II/DAΦNE longitudinal feedback*

This is also adopted at SPEAR, Bessy II, and PLS [22, 16, 17]. The A/D and D/A conversions are performed by VXI boards, while the feedback processing is done by DSP boards hosted in a number of VME crates (Fig. 49).

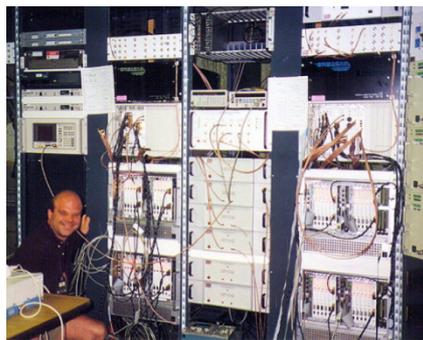

**Fig. 49:** The ALS/PEP-II/DAΦNE longitudinal feedback system

#### 3.4.2.3 *PEP-II transverse feedback*

The digital part, made of two ADCs, a FPGA and a DAC, features a digital delay and integrated diagnostics tools, while the rest of the signal processing is analogue [23, 24].

### 3.4.2.4 KEKB transverse and longitudinal feedback

The digital processing unit, made of discrete digital electronics and banks of memories (Fig. 50), performs a two-tap FIR filter featuring stable beam rejection, phase shift, and delay [25, 26].

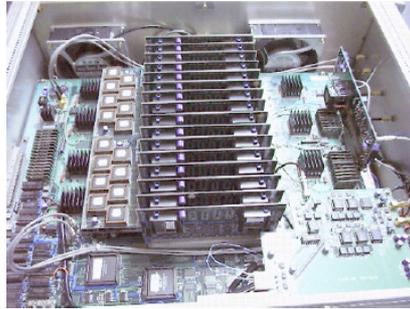

**Fig. 50:** The digital processing board of the KEKB feedback systems

### 3.4.2.5 Elettra/SLS transverse and longitudinal feedback

The digital processing unit is made of a VME crate equipped with one ADC, one DAC and six commercial DSP boards with four microprocessors each (Fig. 51) [27, 28]. Further developments at SLS allowed the direct connection of ADC and DAC boards, exploiting the on-board FPGAs to perform feedback processing [29].

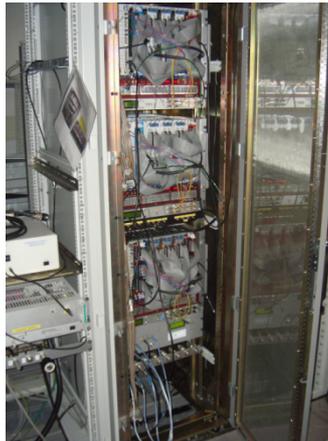

**Fig. 51:** The Elettra horizontal, vertical, and longitudinal feedback processors

### 3.4.2.6 CESR transverse and longitudinal feedback

These employ VME digital processing boards equipped with ADC, DAC, FIFOs and PLDs [30].

### 3.4.2.7 HERA-p longitudinal feedback

This is made of a processing chain with two ADCs (for I and Q components), a FPGA and two DACs [31].

### 3.4.2.8 HLS tranverse feedback

The digital processor consists of two ADCs, one FPGA and two DACs [32].

### 3.4.2.9 SPring-8 transverse feedback

This is also adopted at TLS, KEK-Photon-Factory, and Soleil [33–36]. It employs a fast analogue de-multiplexer that distributes analogue samples to a number of slower ADC–FPGA channels. The correction samples are converted to analogue by one DAC (Fig. 52).

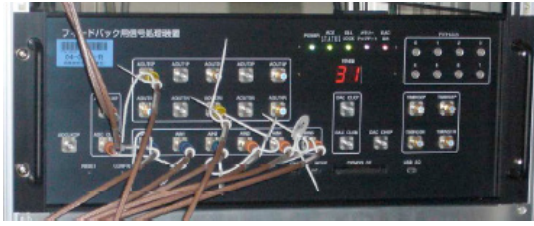

**Fig. 52:** The processing unit of the SPring-8 multi-bunch feedback system

*3.4.2.10 ESRF transverse/longitudinal and Diamond transverse feedback*

These adopt a commercial product called Libera Bunch-by-bunch [20, 37] from Instrumentation Technologies [38] (Fig. 53), which features four ADCs sampling the same analogue signal appropriately delayed, one FPGA and one DAC.

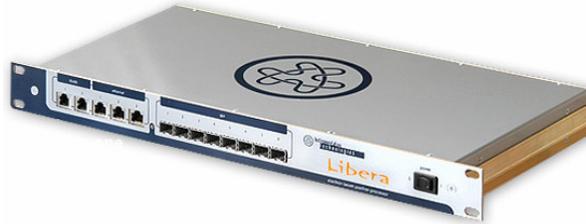

**Fig. 53:** Libera Bunch-by-bunch by Instrumentation Technologies

*3.4.2.11 DAΦNE transverse and KEK-Photon-Factory longitudinal feedback*

These feedback systems [39] rely on a commercial product called iGp from Dimtel [40] (Fig. 54), featuring a ADC–FPGA–DAC chain.

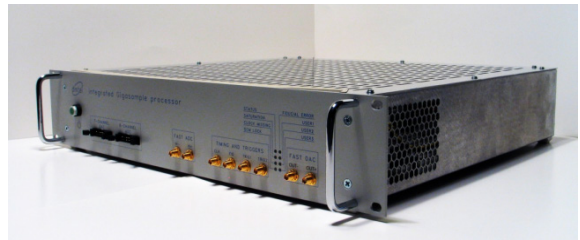

**Fig. 54:** iGp by Dimtel

## 3.5 Amplifier and kicker

The kicker is the feedback actuator. It generates a transverse/longitudinal electromagnetic field that steers the bunches with small 'kicks' as they pass through the kicker. The overall effect is the damping of the betatron/synchrotron oscillations. The power amplifier provides the kicker with the necessary RF power by amplifying the signal from the DAC (or from the modulator in the case of longitudinal feedback).

An important parameter that measures the efficiency of the kicker is the shunt impedance $R$, defined as the ratio between the squared voltage seen by the bunch and twice the power at the kicker input

$$R = \frac{V^2}{2P_{IN}} \ . \qquad (11)$$

Since the shunt impedance depends on the frequency of the input signal, it is normally represented as a function of frequency. We shall use this definition below, when calculating the power necessary to damp coupled-bunch oscillations.

The amplifier and kicker need a bandwidth of at least $f_{rf}/2$. In a transverse feedback, for example, frequencies go from ~DC (all kicks of the same sign) to ~$f_{rf}/2$ (kicks of alternating signs). The bandwidth of the amplifier and kicker must be sufficient to correct each bunch with the appropriate kick without affecting the neighbouring bunches. The amplifier and kicker design has to maximize the kick strength while minimizing the cross-talk between corrections given to adjacent bunches.

Another issue is the group delay that should be constant in the working frequency range, namely, the phase response should be linear. If this is not the case, the feedback efficiency is reduced when damping some multi-bunch modes and, under given conditions, the feedback can even become positive. Figure 55 shows an example of amplitude and phase response of an amplifier. Where the phase is 180°, the multi-bunch mode corresponding to that frequency is excited.

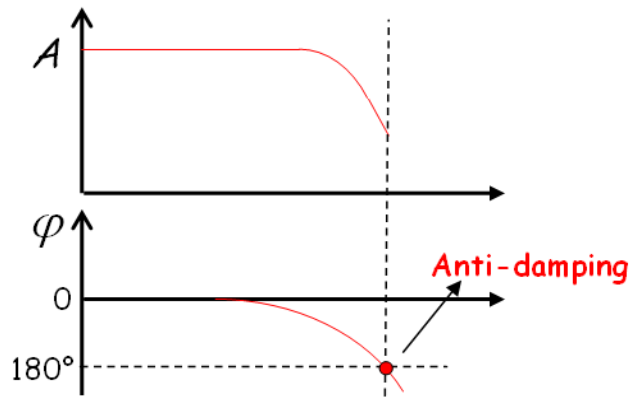

**Fig. 55:** An example of amplifier amplitude and phase response where, for a given frequency, the feedback performs anti-damping

### 3.5.1 *Transverse feedback*

A stripline geometry is usually employed for the transverse feedback kicker. The amplifier and kicker work in baseband, namely in the frequency range from ~DC to ~$f_{rf}/2$. Figure 56 shows an example of a transverse feedback back-end adopting two power amplifiers feeding the downstream ports of two striplines in counter-phase. The upstream ports are connected to 50 Ω power loads. Power low-pass filters can optionally be included to protect the amplifiers from peak voltages picked up from the beam by the kicker.

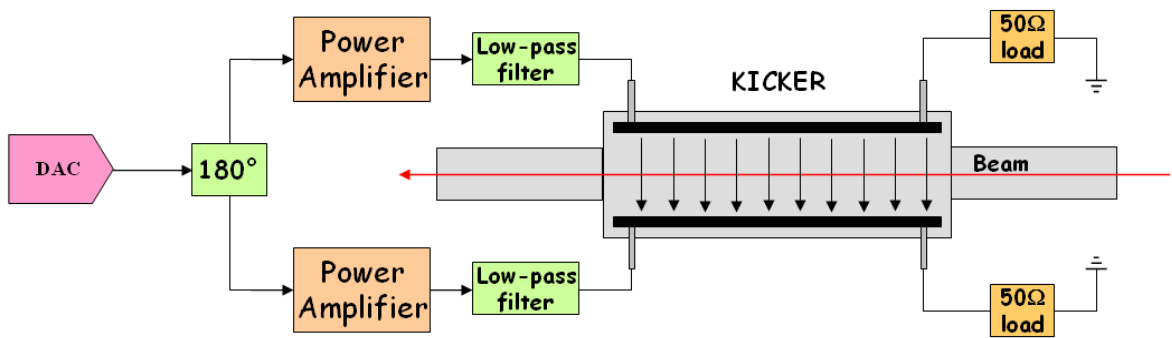

**Fig. 56:** An example of a transverse feedback back-end using two power amplifiers to feed the kicker

The design of the Elettra/SLS transverse kickers [41] is depicted in Fig. 57(a) together with a picture of the kickers installed on the Elettra vacuum chamber (Fig. 57(b)), while the shunt impedance of the same kickers as a function of frequency is plotted in Fig. 58.

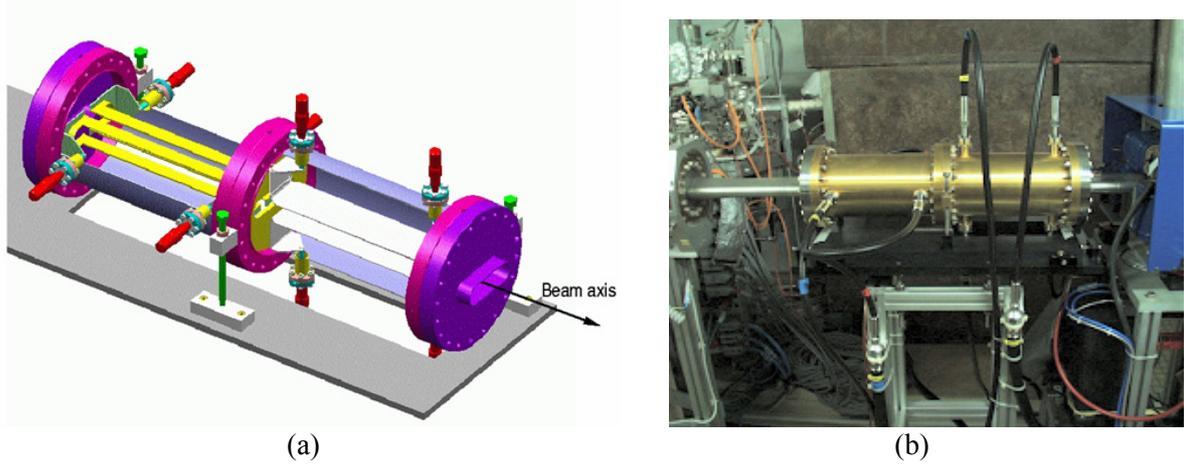

(a) (b)
**Fig. 57:** (a) The Elettra/SLS horizontal and vertical kickers; (b) kickers installed on the Elettra vacuum chamber

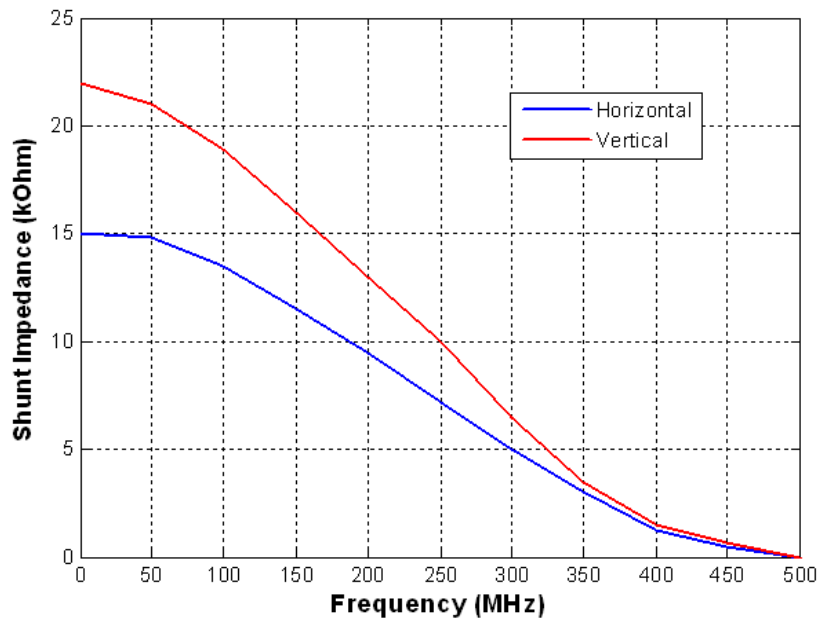

**Fig. 58:** Shunt impedance of the Elettra/SLS transverse kickers

### *3.5.2 Longitudinal feedback*

For the longitudinal kicker, a 'cavity-like' structure is usually preferred because of the higher shunt impedance and smaller size. As in the case of the transverse kicker, the operating frequency range of the longitudinal kicker is $f_{rf}/2$ wide, but it is usually placed on one side of a multiple of $f_{rf}$ (e.g. from $3f_{rf}$ to $3f_{rf} + f_{rf}/2$).

The baseband signal from the DAC has to be translated in frequency or, in other words, 'modulated' in order to overlap with the kicker shunt impedance (Fig. 59). Single SideBand (SSB) amplitude modulation or other techniques such as, for example, Quadrature Phase Shift Keying (QPSK) modulation can be adopted. The modulated signal is then amplified and sent to the kicker. In this case a circulator can be employed to protect the amplifier from reverse power (Fig. 60) [42]. The Elettra/SLS longitudinal kicker is illustrated (Fig. 61) [43].

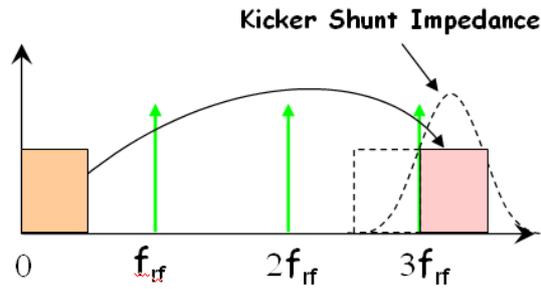

**Fig. 59:** The baseband signal is modulated to overlap with the kicker shunt impedance

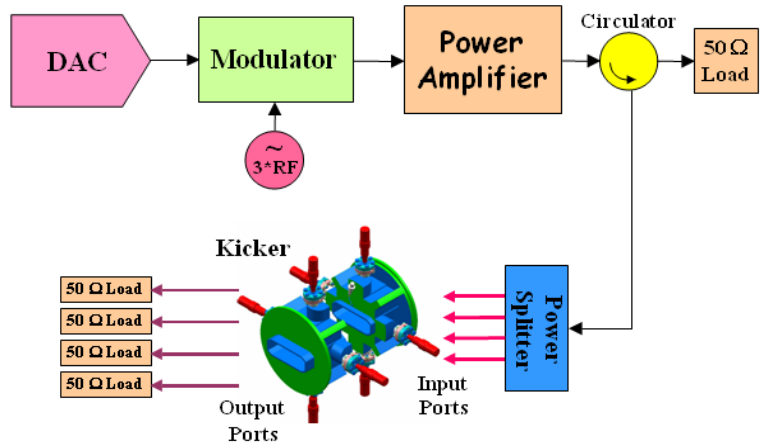

**Fig. 60:** Block diagram of a typical longitudinal feedback back-end

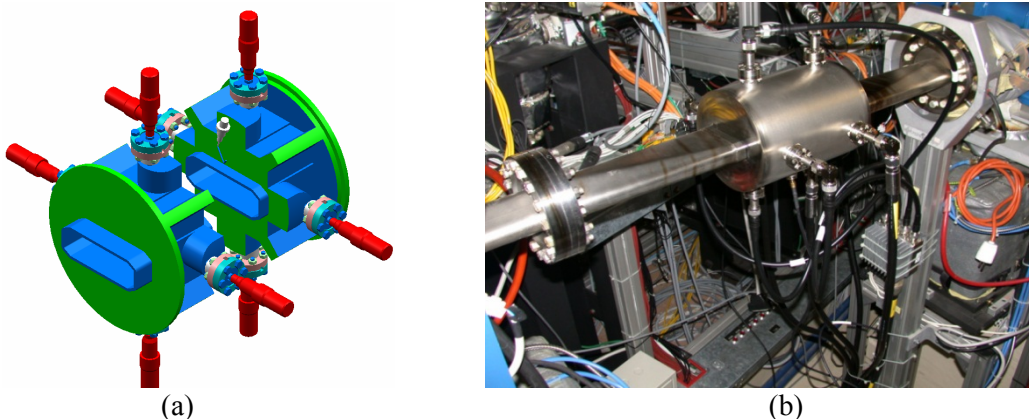

(a)                         (b)
**Fig. 61:** (a) The Elettra/SLS longitudinal kicker; (b) kicker installation on the Elettra vacuum chamber

### 3.6 Control system integration

Each component of the feedback system that needs to be configured and adjusted should include an interface to the accelerator control system (Fig. 62). There should be the ability to perform any operation remotely to facilitate system commissioning and the optimization of its performance. Moreover, as we shall see below, it is crucial to have an effective data acquisition channel to provide fast transfer of large amounts of data for analysis of feedback performance and for beam dynamics studies. It is also preferable to have direct access to the feedback system from a numerical computing environment and/or a script language like MATLAB [44] or similar products (Octave, Scilab, Python, IGOR Pro, IDL, etc.) for quick development of measurement procedures using scripts as well as for data analysis and visualization.

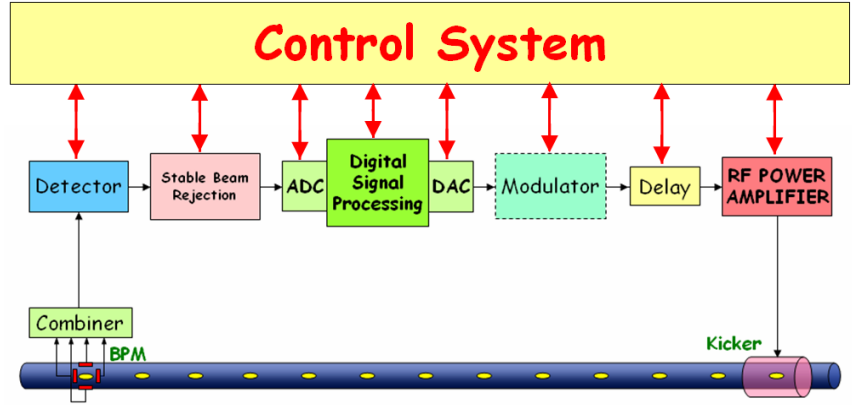

**Fig. 62:** Integration of the feedback system components into the accelerator control system

## 4 RF power requirements

In this section we determine the formulae to calculate the RF power required to damp coupled-bunch instabilities.

### 4.1 Transverse feedback

The transverse motion of a bunch of particles not subject to damping or excitation can be described as a pseudo-harmonic oscillation with amplitude proportional to the square root of the $\beta$ function

$$x(s) = a\sqrt{\beta(s)}\cos\phi(s), \quad \text{where} \quad \phi(s) = \int_0^s \frac{d\bar{s}}{\beta(\bar{s})} \tag{12}$$

where $s$ is the longitudinal coordinate. The derivative of the position, namely the angle of the trajectory, is

$$x' = -\frac{a}{\sqrt{\beta}}\sin\phi + \frac{a\beta'}{2\sqrt{\beta}}\cos\phi, \quad \text{with} \quad \phi' = \frac{1}{\beta} \tag{13}$$

By introducing $\alpha = -\frac{\beta'}{2}$ we can write

$$x' = \frac{a}{\sqrt{\beta}}\sqrt{1+\alpha^2}\,\sin(\varphi + \arctan\alpha) \tag{14}$$

Let us consider now a kicker placed at coordinate $s_k$; the kicker electromagnetic field deflects the bunch trajectory, which varies its angle by $k$. As a consequence the bunch starts another oscillation

$$x_1 = a_1\sqrt{\beta}\cos\varphi_1 \tag{15}$$

which must satisfy the following constraints

$$\begin{cases} x(s_k) = x_1(s_k) \\ x'(s_k) = x_1'(s_k) + k \end{cases} \tag{16}$$

By introducing
$$A = a\sqrt{\beta}, \quad A_1 = a_1\sqrt{\beta} \tag{17}$$

Eq. (16) becomes a two-equation, two-unknown-variables system:

$$\begin{cases} A\cos\varphi = A_1 \cos\varphi_1 \\ A\frac{\sqrt{1+\alpha^2}}{\beta}\sin(\varphi+\text{arctg}(\alpha)) = A_1\frac{\sqrt{1+\alpha^2}}{\beta}\sin(\varphi_1+\text{arctg}(\alpha)) + k. \end{cases} \tag{18}$$

The solution of the system gives amplitude and phase of the new oscillation

$$\begin{cases} A_1 = \sqrt{(A\sin\phi - k\beta)^2 + A^2\cos^2\phi} \\ \phi_1 = \arccos\left(\frac{A}{A_1}\cos\phi\right). \end{cases} \tag{19}$$

From the first equation, if the kick is small ($k \ll A/\beta$) then
$$\frac{\Delta A}{A} = \frac{A - A_1}{A} \cong \frac{\beta}{A} k \sin\phi. \tag{20}$$

From the same equation it can be seen that in the linear feedback case, i.e. when the turn-by-turn kick signal is a sampled sinusoid with amplitude proportional to the bunch oscillation amplitude, in order to maximize the damping rate the kick signal must be in phase with $\sin\varphi$

$$k = g\frac{A}{\beta}\sin\varphi \quad \text{with} \quad 0 < g < 1, \tag{21}$$

that is, in quadrature with the bunch oscillation (Eq. (12)).

The optimal gain $g_{opt}$ (which must in any case be <1) is determined by the maximum kick value $k_{max}$ that the kicker is able to generate. If we want the feedback to work in a linear regime, i.e. not in saturation, the feedback gain must be set so that $k_{max}$ is generated when the oscillation amplitude $A$ at the kicker location is maximum: $g_{opt} = \frac{k_{max}}{A_{max}}\beta$. Therefore

$$k = \frac{k_{max}}{A_{max}} A \sin\varphi. \tag{22}$$

For small kicks the relative amplitude decrease is
$$\frac{\Delta A}{A} \cong \frac{k_{max}}{A_{max}}\beta\sin^2\varphi \tag{23}$$

and its average is
$$\left\langle \frac{\Delta A}{A} \right\rangle \cong \frac{\beta\, k_{max}}{2\, A_{max}}. \tag{24}$$

The average relative decrease is therefore constant, which means that, on average, the amplitude decrease is exponential with time constant $\tau$ (damping time) given by

$$\frac{1}{\tau} = \left\langle \frac{\Delta A}{A} \right\rangle \frac{1}{T_0} = \frac{\beta\, k_{max}}{2\, A_{max}\, T_0}, \tag{25}$$

where $T_0$ is the revolution period. By referring to the oscillation amplitude at the BPM location

$$\frac{1}{\tau} = \frac{k_{max}}{2 T_0 A_{Bmax}} \sqrt{\beta_K \beta_B} \tag{26}$$

where $A_{Bmax}$ is the maximum oscillation amplitude at the BPM location.

In the case of relativistic particles, the change of the transverse momentum $p$ of the bunch passing through the kicker can be expressed by

$$\Delta p = \frac{e}{c} V_\perp ,$$

where

$$V_\perp = \int_0^L (\bar{E} + c \times \bar{B})_\perp \, dz$$

is the kick voltage, $e$ is the electron charge, $c$ is the speed of light, $\bar{E}$ and $\bar{B}$ are the fields in the kicker and $L$ is the length of the kicker. Here, $p$ can be written as $\frac{E_0}{c}$, where $E_0$ is the beam energy and $V_\perp$ can be derived from the definition of kicker shunt impedance

$$R_k = \frac{V_\perp^2}{2 P_k} .$$

The maximum deflection angle in the kicker is given by

$$k_{max} = \frac{\Delta p}{p} = e \frac{V_\perp}{E_0} = \left(\frac{e}{E_0}\right) \sqrt{2 P_k R_k} . \tag{27}$$

From the previous equations we obtain

$$P_K = \frac{2}{R_K} \left(\frac{E_0}{e}\right)^2 \left(\frac{T_0}{\tau}\right)^2 \left(\frac{A_{Bmax}}{\sqrt{\beta_B}}\right)^2 . \tag{28}$$

Given a free bunch oscillation with amplitude at the BPM location $A_{Bmax}$, Eq. (28) allows us to calculate the power necessary to damp this oscillation with time constant $\tau$. In the case of excited oscillations, in order to damp the instability, the feedback damping time constant must be less than the instability growth time-constant (see Section 1.3.3).

It has to be noted that this equation is only valid if all of the system components are ideal and the feedback is perfectly tuned.

The maximum power is supplied when the oscillation amplitude is at its maximum, namely at the time the feedback is switched on. During the damping process, the power decreases with the square of the oscillation amplitude. This leads us to an important conclusion: in order to determine the required power of the RF amplifier, we have to know not only the strength of the multi-bunch instability we want to damp, but also its maximum amplitude. For the same reason it is preferable to switch on the feedback when the oscillation is small: if we keep the feedback gain high (a small oscillation corresponds to the entire dynamic range of DAC/amplifier) it will be easier to catch and keep it damped.

The following is an example of the calculation of the required power for the Elettra transverse feedback system. The parameters are $R_k = 15$ kΩ (average value), $E_B/e = 2$ GeV, $T_0 = 864$ ns, $\tau = 120$ μs, $\beta_{B\ H,V} = 5.2, 8.9$ m, $\beta_{K\ H,V} = 6.5, 7.5$ m.

The plots in Fig. 63 show the required power in the horizontal and vertical planes as a function of the initial oscillation amplitude.

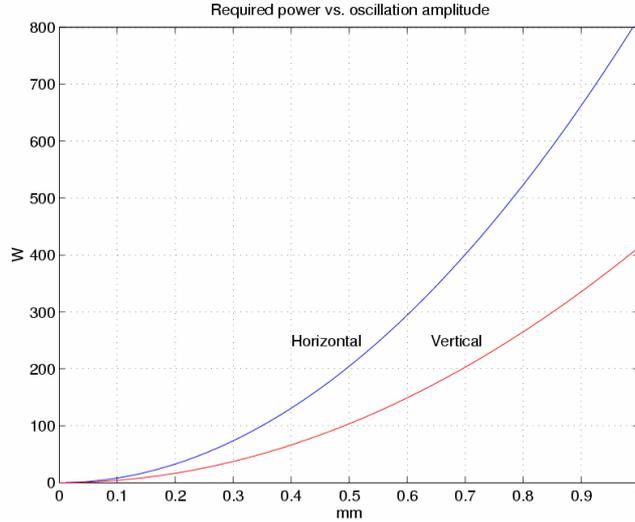

**Fig. 63:** Power required to damp a transverse instability as a function of the initial oscillation amplitude calculated for the Elettra storage ring

### 4.2 Longitudinal feedback

A similar procedure can be adopted to calculate the power required for longitudinal feedback. The relationship between the energy oscillation amplitude $\varepsilon$ and the phase oscillation amplitude $\varphi$ of a longitudinal synchrotron oscillation is given by

$$\varepsilon = \frac{E_0\, \omega_s}{2\pi f_{\rm rf}\, \alpha} \phi , \tag{29}$$

where $\omega_s$ is the synchrotron frequency, $\alpha$ the momentum compaction factor, and $f_{\rm rf}$ the RF frequency. As in the transverse case, the damping time constant $\tau$ is given by the equation

$$\frac{1}{\tau} = \frac{eV_{\max}}{2\varepsilon_{\max} T_0} , \tag{30}$$

where $V_{\max}$ is the maximum kick voltage and $\varepsilon_{\max}$ is the maximum energy oscillation amplitude.

By introducing the longitudinal kicker shunt impedance $R_k$,

$$R_k = \frac{V^2}{2 P_k} , \tag{31}$$

the power required to damp a longitudinal oscillation is given by

$$P_k = \frac{2}{R_k} \left( \frac{\omega_s}{\omega_0} \frac{E_0}{e} \frac{\phi_{\max}}{\alpha\, f_{\rm rf}\, \tau} \right)^2 . \tag{32}$$

## 5 Digital signal processing

As mentioned in Section 3.4, the digital processing in a bunch-by-bunch feedback is performed by *M* separated channels, each dedicated to one bunch (Fig. 64).

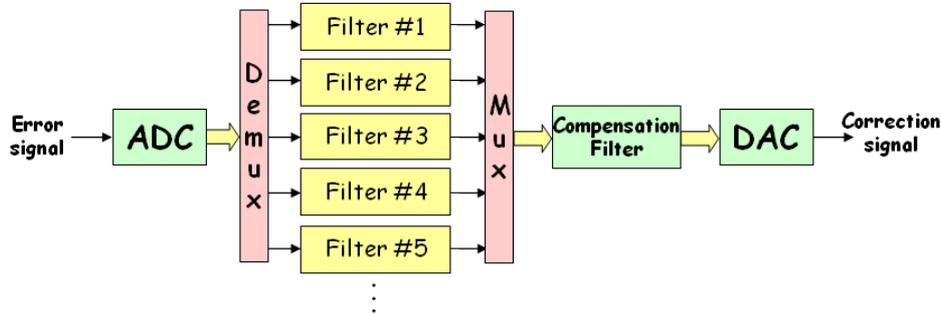

**Fig. 64:** Block diagram of the digital processor

In order to damp the bunch oscillations, the turn-by-turn kick signal must be the derivative of the bunch position at the kicker, namely, for a given oscillation frequency, a $\pi/2$ phase-shifted signal must be generated. In determining the real phase shift to perform in each channel, the phase advance between BPM and kicker must be taken into account as well as any additional delay due to the feedback latency (multiple of one machine revolution period). Moreover, any residual constant offset (stable beam component) must be rejected from the bunch signal to avoid DAC saturation.

These basic tasks can be accomplished by digitally filtering the streams of turn-by-turn samples in each channel, namely calculating the present correction sample on the basis of the past position samples.

Additionally, to compensate for the not-ideal characteristics of an amplifier and kicker, a digital filter can be put in the full-rate data path before the DAC.

### 5.1 Digital filter design

Digital filters can be implemented with finite impulse response (FIR) or infinite impulse response (IIR) structures. In general, IIR filters offer better performance than FIR filters. The disadvantage is a more difficult design and an increased complexity of the filter implementation.

Various techniques are used to design digital filters, including frequency domain and model-based design. In the following sections we show some examples of digital filters actually adopted in bunch-by-bunch feedback systems.

#### 5.1.1 Three-tap FIR filter

The minimum requirements for each of the *M* digital filters are:

i) DC rejection (for a FIR filter this means that the sum of the coefficients must be zero);
ii) given amplitude response at the tune frequency;
iii) given phase response at the tune frequency.

A simple three-tap FIR filter can fulfil these requirements. Since it is a very simple filter, the coefficients can be calculated analytically. Let us consider the Z-transform of the filter response, namely the filter transfer function *H(z)*; in order to have zero amplitude response at DC, *H(z)* must have a 'zero' in $z = 1$. Another zero in $z = c$ is needed in order to have the required phase response $\alpha$ at the tune frequency $\omega$:

$$H(z) = k(1 - z^{-1})(1 - cz^{-1}) \quad . \tag{33}$$

The parameter $c$ can be calculated analytically

$$H(z) = k(1-(1+c)z^{-1} + c\,z^{-2}) \quad z = e^{j\omega}$$
$$H(\omega) = k(1-(1+c)e^{-j\omega} + c\,e^{-2j\omega})$$
$$e^{-j\omega} = \cos\omega - j\sin\omega, \quad \alpha = \text{ang}(H(\omega)) \quad (34)$$
$$\text{tg}(\alpha) = \frac{c(\sin(\omega) - \sin(2\omega)) + \sin(\omega)}{c(\cos(2\omega) - \cos(\omega)) + 1 - \cos(\omega)}$$
$$c = \frac{\text{tg}(\alpha)(1-\cos(\omega)) - \sin(\omega)}{(\sin(\omega) - \sin(2\omega)) - \text{tg}(\alpha)(\cos(2\omega) - \cos(\omega))}.$$

Here $k$ is determined given the required amplitude response at tune $|H(\omega)|$:

$$k = \frac{|H(\omega)|}{\sqrt{(1-(1+c)\cos(\omega) + c\cos(2\omega))^2 + ((1+c)\sin(\omega) - c\sin(2\omega))^2}}. \quad (35)$$

As an example: $\omega/2\pi = 0.2$, $|H(\omega)| = 0.8$, $\alpha = 222°$.

From the equations we can calculate $c = -0.63$ and $k = 0.14$. The transfer function of the filter is therefore $H(z) = -0.63 + 0.49\,z^{-1} + 0.14\,z^{-2}$. The filter response is depicted in Fig. 65.

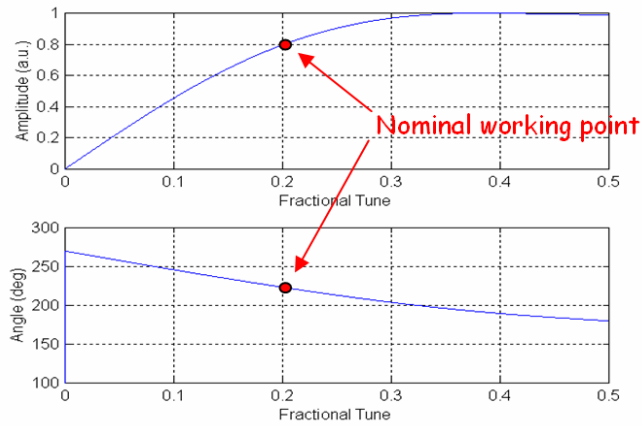

**Fig. 65:** Amplitude and phase response of a three-tap FIR filter

### *5.1.2 Five-tap FIR filter*

With a longer FIR filter more degrees of freedom are available in the design, and additional features can be added to the filter. The following is an example of a five-tap FIR filter currently used at Elettra for transverse feedback [28].

The tune of an accelerator can significantly change during machine operations; the filter can be designed in order to guarantee the same feedback efficiency in a given frequency range. In this case the requirement is to have an optimal phase response.

In this example the feedback delay is four machine turns, which must be taken into account when calculating the filter phase response. When the tune frequency increases, the phase of the filter must increase as well; this means that the phase response must have a positive slope around the working point.

The filter design can be made using the MATLAB function *invfreqz()*. This function calculates the FIR filter coefficients that best fit the required frequency response using the least-squares method. The desired response is specified by defining amplitude and phase at three different frequencies: the origin and two frequencies, $f_1$ and $f_2$, chosen in the proximity of the nominal tune (Fig. 66, lower plot).

As we can see in the upper plot, this is achieved at the expense of a bad amplitude response, which features a minimum at the tune while the maximum is at a different frequency.

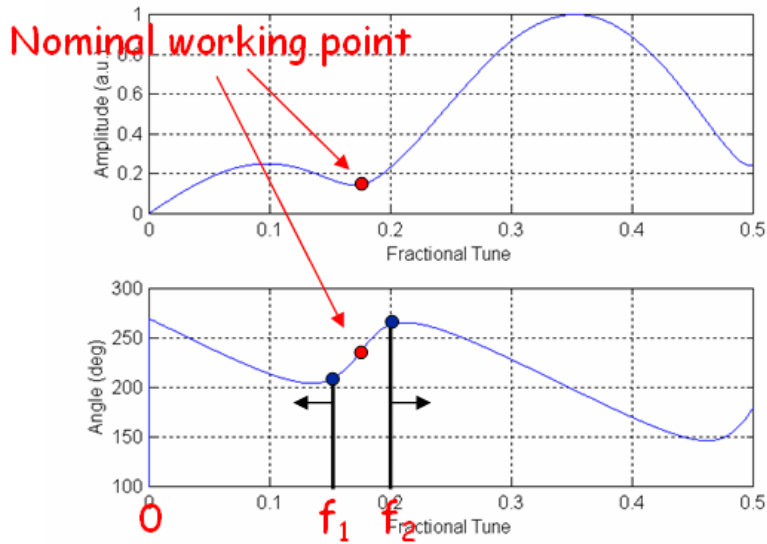

**Fig. 66:** Amplitude and phase response of a five-tap FIR filter

### 5.1.3 *Selective FIR filter*

A filter often employed in longitudinal feedback systems is the selective FIR filter, whose impulse response (the filter coefficients) is a sampled sinusoid with frequency equal to the synchrotron tune (see Fig. 67) [45]. As we can see in Fig. 68, the filter amplitude response has a maximum at the tune frequency and linear phase. The more filter coefficients we use the more selective the filter.

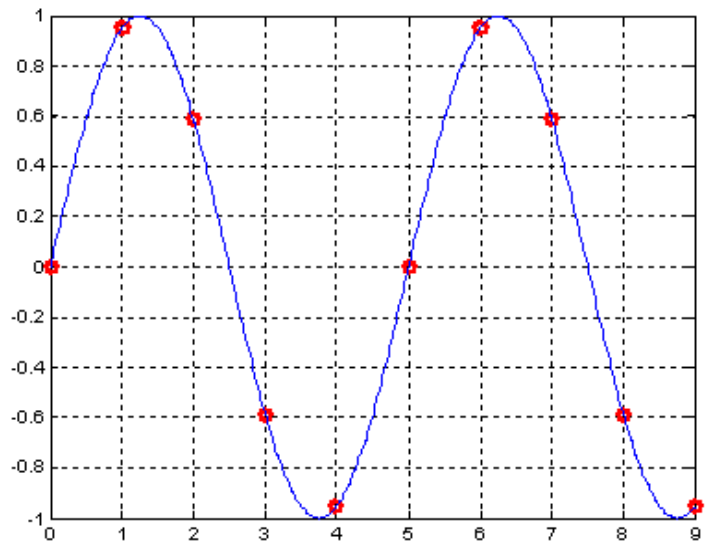

**Fig. 67:** Impulse response of a selective FIR filter (dots)

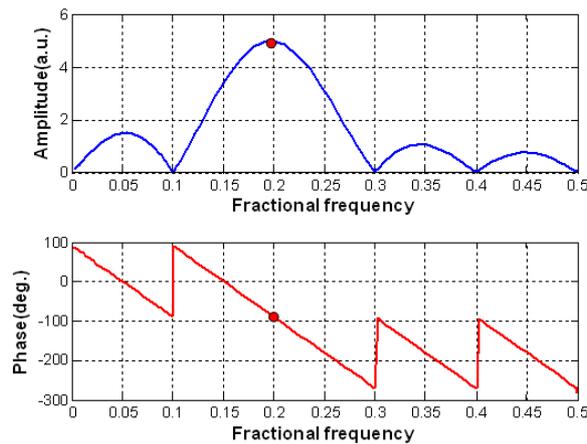

**Fig. 68:** Amplitude and phase response of a selective FIR filter

### *5.1.4 Advanced filter design*

More sophisticated techniques using longer FIR or IIR filters enable a variety of additional features that exploit the potential of digital signal processing.

It is possible, for instance, to enlarge the working frequency range (as seen in Section 5.1.2) with no degradation of the amplitude response. Moreover, the filter selectivity can be enhanced to better reject unwanted frequency components (noise) or to minimize the amplitude response at frequencies that must not be fed back. An example of this design technique is the Time Domain Least Square Fitting (TDLSF) described in Ref. [33].

Another interesting possibility is to stabilize different tune frequencies simultaneously by designing a filter with two separate working points. This is required, for example, when horizontal and vertical as well as dipole and quadrupole instabilities have to be addressed by the same feedback system [35].

Advanced design techniques can also be employed to improve the robustness of the feedback under parametric changes of accelerator or feedback components (e.g. optimal control, robust control, etc.) [46].

### 5.2 Down-sampling

The synchrotron frequency is usually much lower than the revolution frequency, which means that one complete synchrotron oscillation is accomplished in many machine turns. In longitudinal feedback systems, in order to be able to properly filter the bunch signal, down-sampling is usually carried out [47]; one out of $D$ bunch samples is used, where $D$ is the down-sampling factor (Fig. 69).

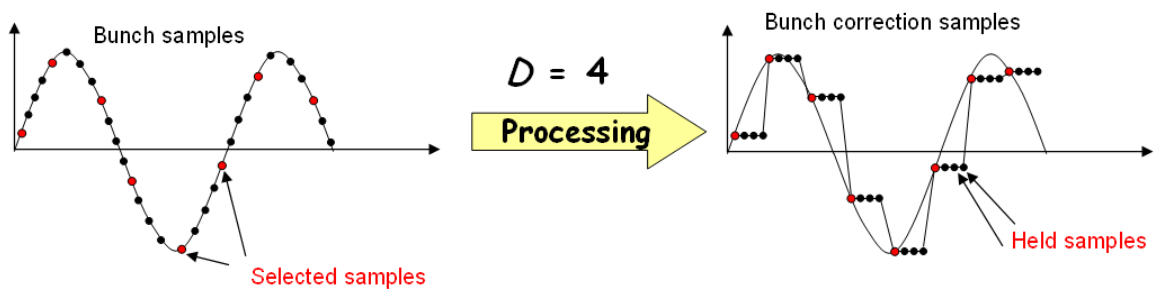

**Fig. 69:** Down-sampling of the bunch position signal and reconstruction by the hold buffer

Filtering is performed over the down-sampled digital signal, and the filter design is done in the down-sampled frequency domain (the original one enlarged by the down-sampling factor $D$). After processing, the turn-by-turn correction signal is reconstructed by means of a hold buffer that keeps each calculated correction value for *D* turns, as shown in Fig. 69.

The reduced data rate allows for more time to be available to perform the filter calculations, and more complex filters can therefore be implemented.

# 6   Integrated diagnostic tools

A feedback system can implement a number of powerful diagnostic tools that are useful for both commissioning and tuning of the feedback system as well as for machine physics studies. The following are some of the tools normally implemented in digital feedback systems:

i)   ADC data recording (see (1) in Fig. 70): acquisition and recording, in parallel with feedback operation, of a large number of samples for off-line data analysis;

ii)  modification of filter parameters 'on the fly' with the required timing and even individually for each bunch (see (2) in Fig. 70): useful for switching feedback on/off, generation of grown/damped transients, optimization of feedback performance, etc.;

iii) injection of externally generated digital samples (see (3) in Fig. 70): for the excitation of single/multi-bunches.

These tools are usually managed by an additional controller with a fast interface to the accelerator control system.

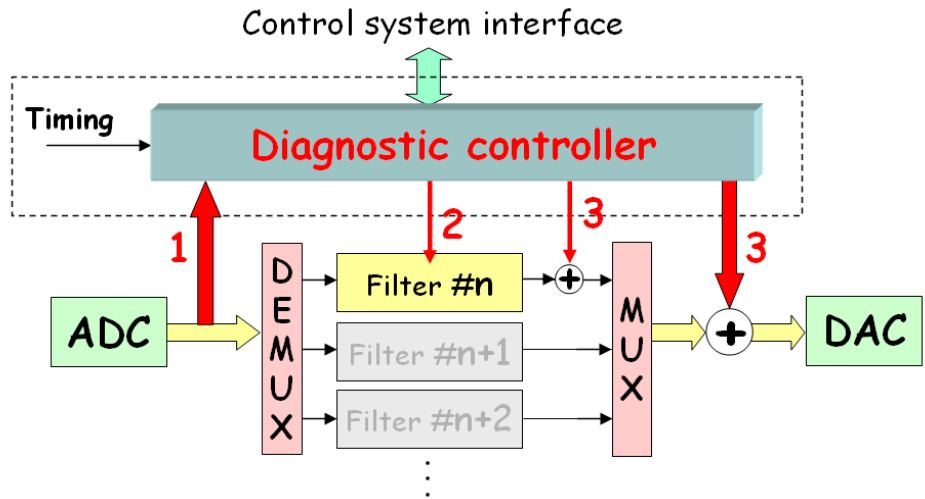

**Fig. 70:** Integrated diagnostics tools of a digital feedback system

Below we give some examples of measurements carried out using the diagnostics tools.

## 6.1   Acquisition and recording of ADC data

Figure 71 depicts the amplitude spectrum of a vertical bunch-by-bunch signal sampled at $f_{rf}$ = 500 MHz. The spectrum is calculated with the FFT algorithm. At the moment of acquisition the beam was vertically unstable with some multi-bunch modes excited around the baseband frequency of 5 MHz.

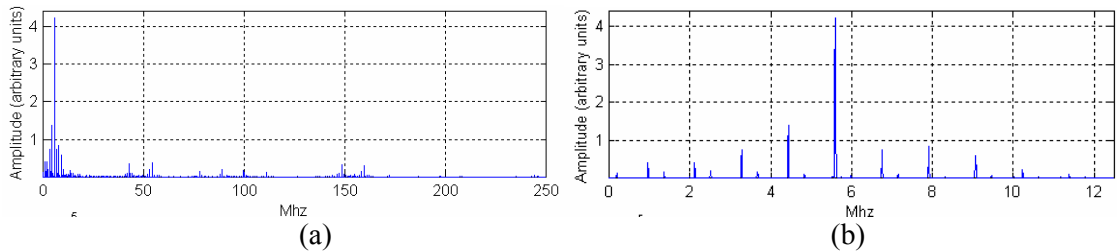

**Fig. 71:** Amplitude spectrum of bunch-by-bunch data of (a) an unstable beam; (b) a zoom view showing the unstable sidebands

### 6.2 Excitation of individual bunches

To produce an excitation, the feedback loop is switched off for one or more selected bunches (for example, by setting to zero the filter coefficients) and an excitation signal is injected in place of the bunch correction signal (Fig. 72). Possible excitation signals are white or pink noise and sinusoids.

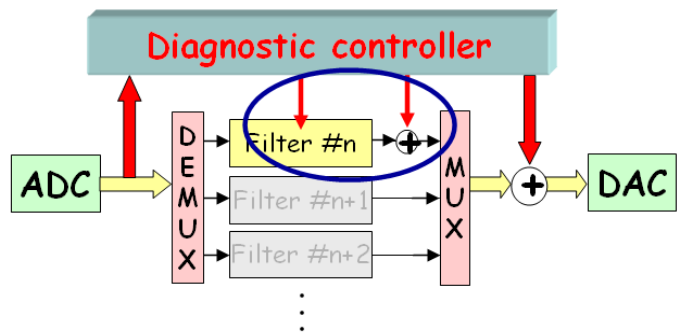

**Fig. 72:** Excitation of individual bunches through injection of appropriate signals

In the example in Fig. 73, two selected bunches are vertically excited with pink noise in a range of frequencies around the tune, while feedback is applied to the other bunches. The upper plot shows the excitation of the two bunches. The spectrum of one of the excited bunches reveals a peak at the tune frequency (lower plot). This technique can be used to measure the betatron tune of individual bunches with almost no deterioration of the beam quality.

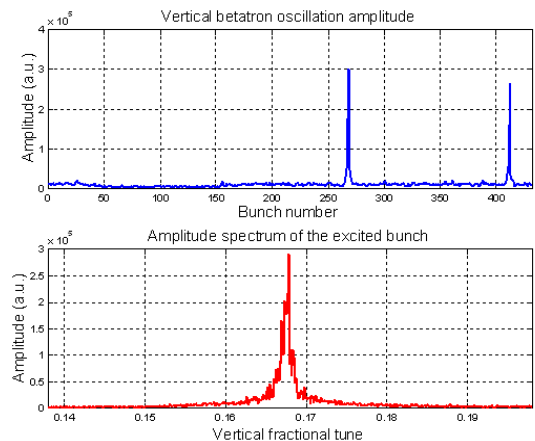

**Fig. 73:** Excitation of individual bunches by injection of appropriate digital signals

Excitation of individual bunches can also be done with feedback on by setting filter coefficients with the sign inverted. This produces positive feedback and anti-damping of the oscillations.

## 6.3 Multi-bunch excitation

Interesting measurements can be done by injecting pre-defined signals into the output of the digital processor (Fig. 74). Depending on the particular excitation signal and the filter settings, several experiments can be carried out.

By injecting a sinusoid at a given frequency, for example, the corresponding beam multi-bunch mode can be excited in order to test the performance of the feedback in damping that mode. If we inject an appropriate signal and record the ADC data when filter coefficients are set to zero, the beam transfer function can be calculated. By doing the same but with filter coefficients set to the nominal values, the closed loop transfer function can be determined.

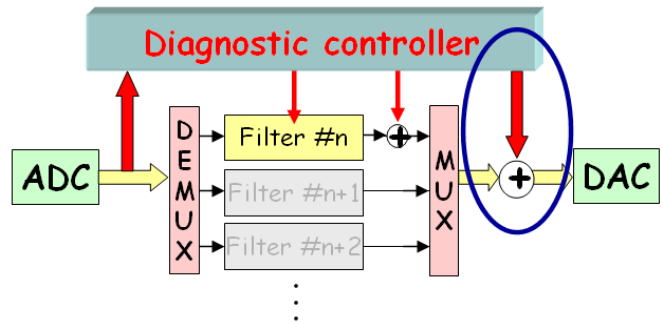

**Fig. 74:** Multi-bunch excitation through injection of appropriate digital signals

## 6.4 Transient generation

A powerful diagnostic application is the generation of transients. Transients can be generated by changing the filter coefficients according to a predefined timing and by concurrently recording the oscillations of the bunches (Fig. 75).

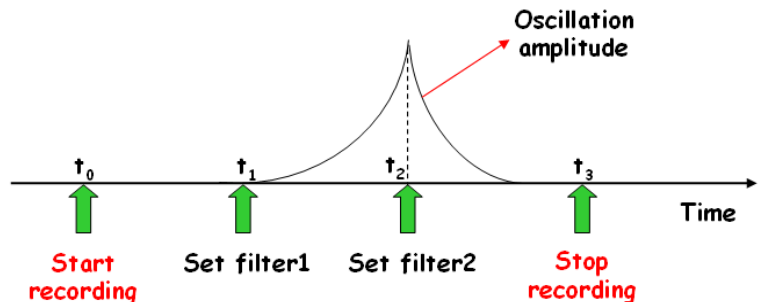

**Fig. 75:** Transient generation by changing the filter coefficients and recording the oscillations

The following are some examples of transients that can be used to measure damping times and growth rates by exponential fitting of the oscillation amplitude.

Let us suppose we have an excited multi-bunch oscillation that has reached equilibrium, i.e. constant amplitude: we can switch on the feedback and record the damping transient (Fig. 76(a)). If the feedback is already on, we can switch it off and on again after a while and generate a grow/damp transient (Fig. 76(b)).

Transients can also be generated by artificially exciting a stable beam with a proper setting of the filter coefficients in order to realize a positive feedback (anti-damping). The feedback is then switched off letting the oscillation decay by natural damping (Fig. 76(c)).

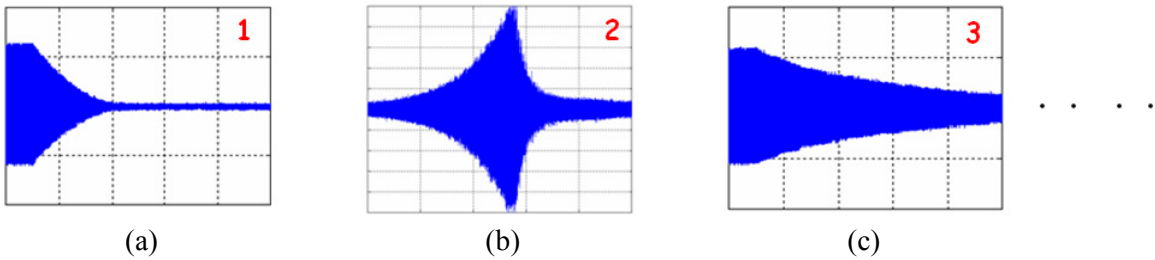

**Fig. 76:** Examples of transient generation

Growing/damped transients can be analysed by means of 3D graphs. Figure 77 is an example where the amplitude of the oscillation vs. time is plotted for every bunch. The absence of oscillation for some bunches is because the corresponding buckets are not filled with particles.

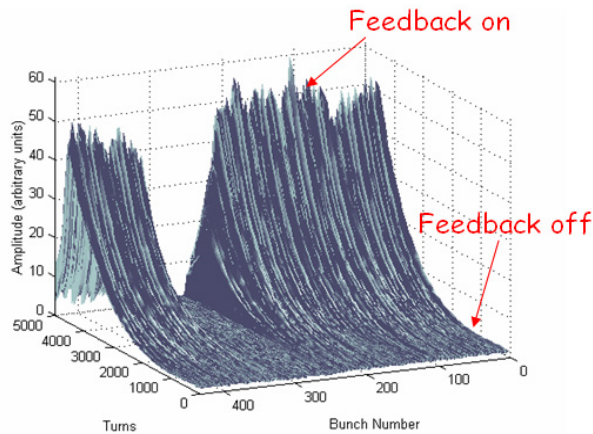

**Fig. 77:** 3D graph showing the evolution of the oscillation amplitude during a growing/damped transient

Transients can also be analysed by displaying the multi-bunch spectrum vs. time to study the evolution of individual coupled-bunch modes (Fig. 78).

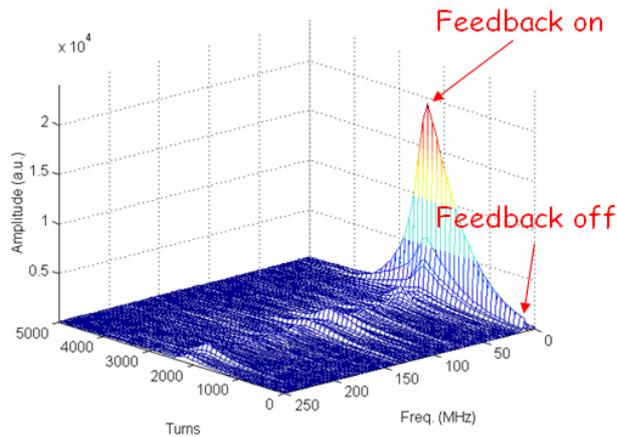

**Fig. 78:** Evolution of coupled-bunch unstable modes during a growing/damped transient

Transient generation is a helpful tool to tune a feedback system as well as to study coupled-bunch modes and beam dynamics [48–53]. The following list mentions some examples of applications using diagnostic tools and, in particular, transient analysis:

i) measurement of the feedback damping time: can be used to characterize and optimize feedback performance;

ii) measurement of the feedback resistive and reactive response: feedback not perfectly tuned can have a reactive behaviour, namely producing a tune shift when switched on that has to be minimized for optimum performance;

iii) modal analysis: measurement and analysis of the complex eigenvalue of the modes, namely the growth rate (real part of the eigenvalue) and the oscillation frequency (imaginary part of the eigenvalue);

iv) impedance measurement: the analysis of complex eigenvalues and bunch synchronous phases can be used to evaluate the machine impedance;

v) stable modes analysis: coupled-bunch modes below the instability threshold, thus being stable, can be studied to predict their behaviour at higher beam currents;

vi) bunch train studies: the analysis of the behaviour of different bunches along the bunch train can be used to determine the sources of coupled-bunch instabilities;

vii) phase space analysis: the analysis of the phase evolution of unstable coupled-bunch modes is used for beam dynamics studies.

## 7  Observation of feedback effects on beam characteristics

In this section we shall see, with the aid of some pictures, some examples of the main observable effects of feedback action measured using beam instrumentation.

### 7.1  Beam spectrum

When the feedback is conveniently tuned the sidebands corresponding to an unstable coupled-bunch mode are completely suppressed. Figure 79 shows an example of beam spectrum with feedback off and on.

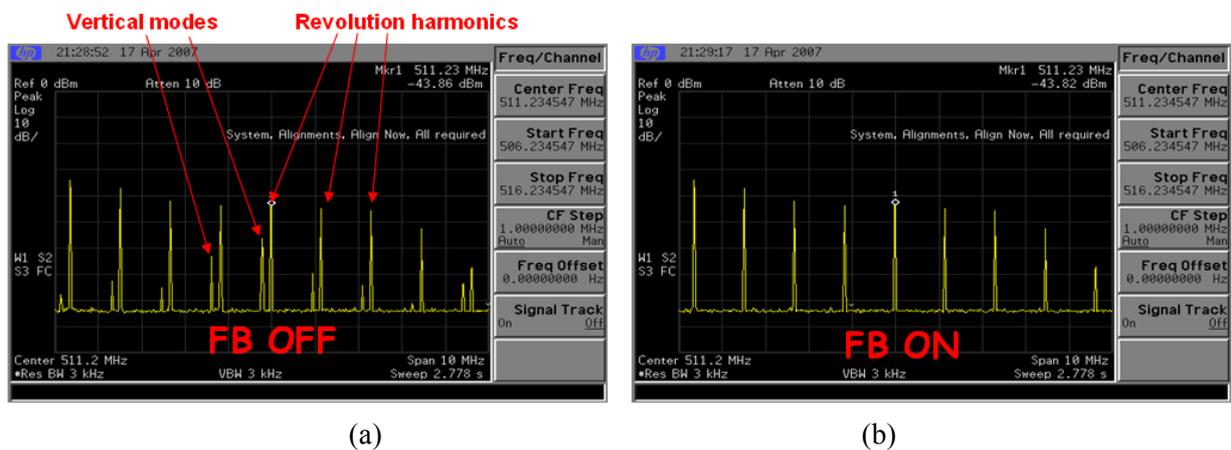

(a)                                                              (b)

**Fig. 79**: Screenshots from a spectrum analyser connected to a stripline pickup. (a) Feedback off; (b) feedback on. The sidebands corresponding to vertical coupled-bunch modes disappear as soon as the transverse feedback is activated.

### 7.2  Beam transverse profile

Images of an electron beam transverse profile can be produced using the synchrotron radiation generated by a bending magnet. As mentioned in Section 1.4.5, the transverse profile blows up in the presence of a transverse instability (Fig. 80(a)). When activated the feedback squeezes the beam, bringing it to its original size (Fig. 80(b)).

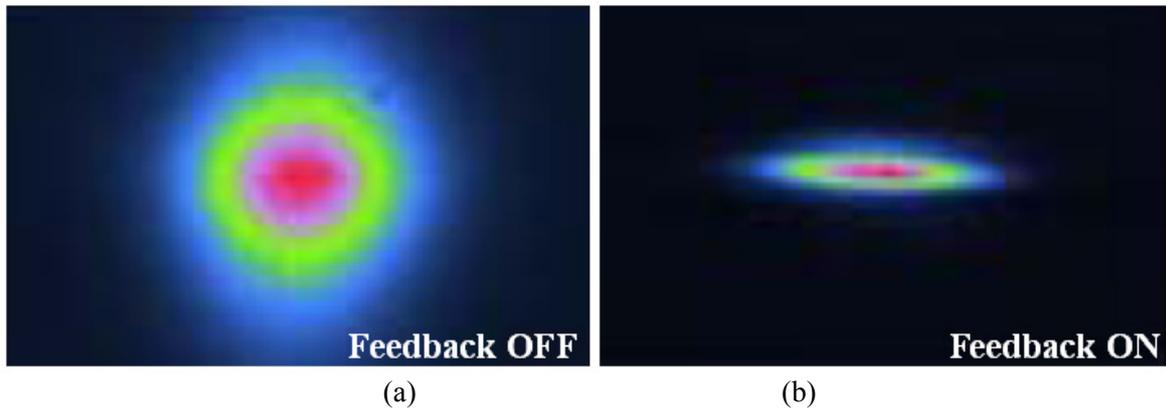

**Fig. 80**: Synchrotron radiation monitor images taken at TLS [34] showing: (a) blow-up of the beam size due to a vertical instability; (b) its reduction when the feedback is switched on.

## 7.3 Streak camera images

Interesting measurements can be carried out by analysing the synchrotron radiation from a bending magnet using a streak camera. In Fig. 81(a) a longitudinal instability is clearly visible. The instability disappears when the feedback is activated (Fig. 81(b)).

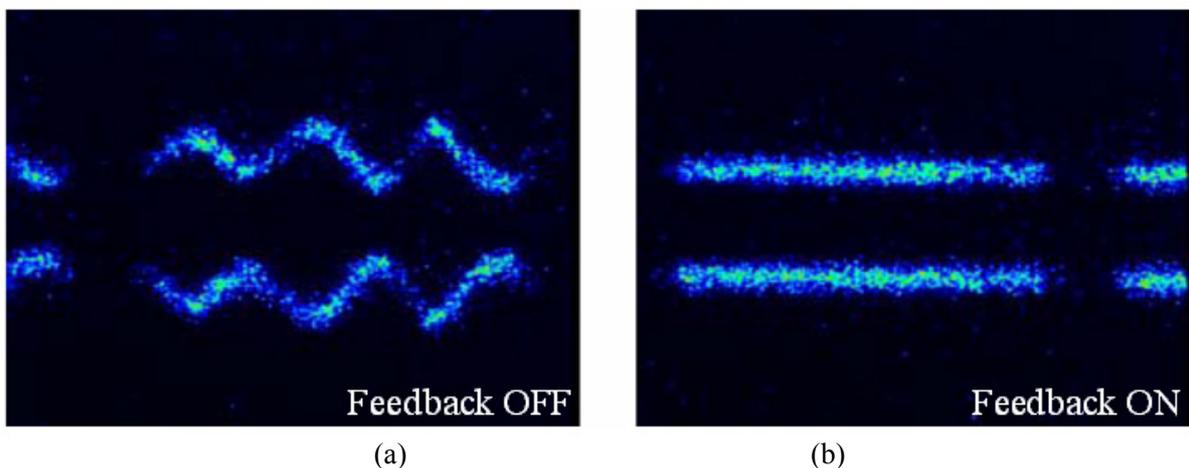

**Fig. 81**: Images of one machine turn taken with a streak camera in 'dual scan mode' at TLS [34]. The horizontal and vertical time spans are 500 and 1.4 ns, respectively. The longitudinal instabilities observable in (a) are suppressed by the feedback in (b).

## 7.4 Photon beam spectra

In a synchrotron light source, the ultimate goal of a feedback system is the improvement of the emitted photon beam quality. This can be evaluated by measuring the energy spectrum of the photons produced by an undulator. As we can observe in the example reported in Fig. 82, the amplitude and shape of the generated harmonics are noticeably improved when vertical instabilities are damped by the feedback.

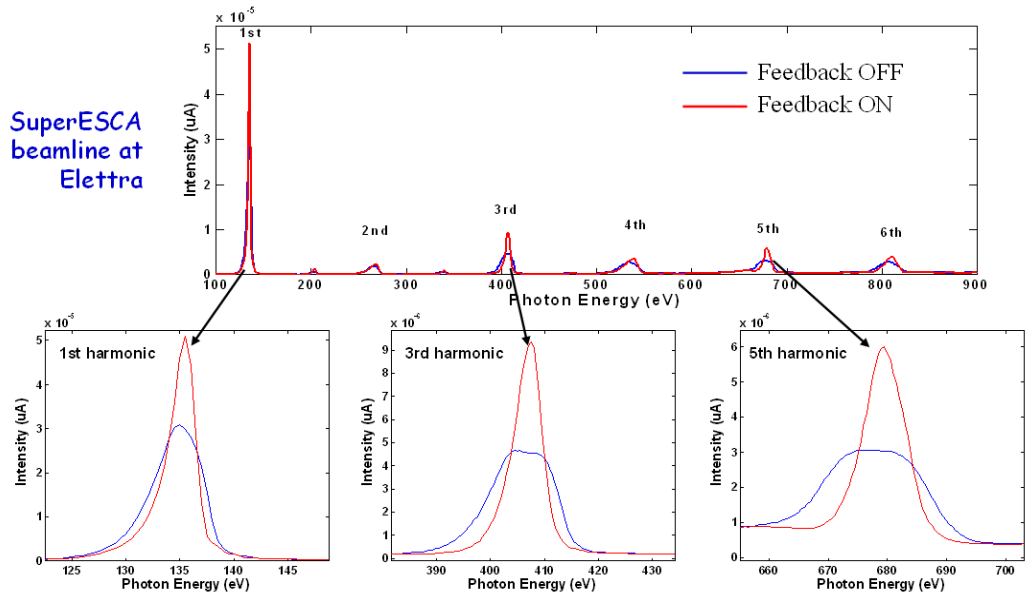

**Fig. 82**: Energy spectra of the photon beam measured on the SuperESCA beamline at Elettra

## 8   Conclusions

Feedback systems are indispensable tools for curing coupled-bunch instabilities in storage rings. Technological advances in digital electronics allow the implementation of digital feedback systems where the control algorithms are executed using programmable devices. The theory of digital signal processing is widely used to design and implement digital filters as well as to analyse data acquired by the systems. In fact, thanks to their capability of exciting the beam in different ways and measuring the resulting beam motion, feedback systems are used not only for closed-loop control but also as powerful diagnostic tools. The possibilities offered by these tools are manifold both for the optimization of feedback performance and for beam dynamics studies.

### Acknowledgements

I would like to thank Dmitry Teytelman, a member of the SLAC feedback team, for many enlightening discussions and useful suggestions on these topics.

### References

[1]   M. Svandrlik *et al.,* The cure of multibunch instabilities in Elettra, Proc. PAC'95, Dallas, 1995.
[2]   E. Courant *et al., Rev. Sci. Instrum.* **37** (1966) 1579.
[3]   M. Zobov *et al.,* Measures to reduce the impedance of parasitic resonant modes in the DAΦNE vacuum chamber, Proc. 14th Advanced IFCA Beam Dynamics Workshop, Frascati, 1997, (Frascati Physics Series Vol. X, 1998), p. 371.
[4]   F. Zimmermann *et al.,* First observations of a 'fast beam ion instability' at the ALS, Proc. PAC'97, Vancouver, 1997.
[5]   R.A. Bosch *et al.,* Suppression of longitudinal coupled bunch instabilities by a passive higher harmonic cavity, Proc. PAC'93, Washington, 1993.
[6]   E. Vogel *et al., Phys. Rev. Spec. Top. – Accel. Beams* **8** (2005) 102801.

[7] G. Lambertson, Control of coupled-bunch instabilities in high-current storage rings, Proc. PAC'91, San Francisco, 1991.

[8] J. Fox *et al.*, Feedback control of coupled-bunch instabilities, Proc. PAC'93, Washington, 1993.

[9] M. Serio *et al.*, Multibunch instabilities and cures, Proc. EPAC'96, Sitges, 1996.

[10] F. Pedersen, *Feedback Systems*, CERN PS/90-49 (AR), 1990.

[11] D Teytelman, Survey of digital feedback systems in high current storage rings, Proc. PAC2003, Portland, 2003.

[12] K. Balewski, Review of feedback systems, Proc. EPAC'98, Stockholm, 1998.

[13] F. Pedersen, Multibunch instabilities, Proc. Frontiers of Particle Beams: Factories with $e^+ e^-$ Rings, Joint US-CERN School on Particle Accelerators, Benalmadena, 1992, pp. 269–292.

[14] F. Pedersen *et al.*, *IEEE Trans. Nucl. Sci.* **NS-24** (1977) 1396.

[15] W. Barry *et al.*, Commissioning of the ALS transverse coupled-bunch feedback system, Proc. PAC'95, Dallas, 1995.

[16] S. Khan *et al.*, BESSY II feedback systems, Proc. PAC'99, New York, 1999.

[17] H.S. Kang *et al.*, Longitudinal and transverse feedback systems for PLS storage ring, Proc. PAC2001, Chicago, 2001.

[18] P. Wesolowsky *et al.*, A vertical multi-bunch feedback system for ANKA, Proc. PAC2005, Knoxville, 2005.

[19] A. Young *et al.*, RF and baseband signal processing in the PEP-II/ALS/DAΦNE longitudinal feedback system, Proc. DIPAC'97, Frascati, 1997.

[20] E. Plouviez *et al.*, Broadband bunch by bunch feedback for the ESRF using a single high resolution and fast sampling FPGA DSP, Proc. EPAC2006, Edinburgh, 2006.

[21] M. Enert *et al.*, Transverse and longitudinal multibunch feedback systems for PETRA, DESY 91–036, 1991.

[22] D. Teytelman *et al.*, Architecture and technology of 500 M sample/s feedback systems for control of coupled-bunch instabilities, Proc. ICALEPCS'99, Trieste, 1999.

[23] J. Byrd *et al.*, Design of the PEP-II transverse coupled-bunch feedback system, Proc. PAC'95, Dallas, 1995.

[24] J. Weber *et al.*, PEP-II transverse feedback electronics upgrade, Proc. PAC2005, Knoxville, 2005.

[25] M. Tobiyama *et al.*, *Phys. Rev. Spec. Top. – Accel. Beams* **3** (2000) 012801.

[26] E. Kikutani *et al.*, Present status of the KEKB bunch feedback systems, Proc. EPAC2002, Paris, 2002.

[27] M Lonza *et al.*, Digital processing electronics for the Elettra transverse multibunch feedback system, Proc. ICALEPCS'99, Trieste, 1999.

[28] D. Bulfone *et al.*, Bunch-by-bunch control of instabilities with the ELETTRA/SLS digital feedback systems, Proc. ICALEPCS 2003, Gyeongju, 2003.

[29] M. Dehler *et al.*, State of the SLS multibunch feedbacks, Proc. APAC2007, Indore, 2007.

[30] J. Sikora *et al.*, Performance of the beam stabilizing feedback systems at CESR, Proc. PAC2001, Chicago, 2001.

[31] J. Randhahn *et al.*, Performance of the new coupled bunch feedback system at HERA-p, Proc. PAC2007, Albuquerque, 2007.

[32] Z.R. Zhou *et al.*, Development of digital transverse bunch-by-bunch feedback system of HLS, Proc. PAC2007, Albuquerque, 2007.

[33] T. Nakamura *et al.*, transverse bunch-by-bunch feedback system for the Spring-8 storage ring, Proc. EPAC2004, Lucerne, 2004.

[34] C.H. Kuo *et al.,* Control of the multi-bunch instabilities at TLS, Proc. APAC2007, Indore, 2007.

[35] W.X. Cheng *et al.,* Single-loop two dimensional transverse feedback for Photon Factory, Proc. EPAC2006, Edinburgh, 2006.

[36] R. Nagaoka *et al.,* Transverse feedback development at SOLEIL, Proc. PAC2007, Albuquerque, 2007.

[37] G. Rehm *et al.,* First tests of the transverse multibunch feedback at Diamond, Proc. DIPAC2007, Venice, 2007.

[38] http://www.i-tech.si/

[39] T. Obina *et al.,* Longitudinal feedback system for the Photon Factory, Proc. PAC2007, Albuquerque, 2007.

[40] http://www.dimtel.com/

[41] M. Dehler *et al.,* Current status of the Elettra/SLS transverse multi-bunch feedback, Proc. EPAC2000, Vienna, 2000.

[42] A. Ghigo *et al.,* Kickers and power amplifiers for the DAΦNE bunch-by-bunch longitudinal feedback system, Proc. EPAC'96, Sitges, 1996.

[43] M. Dehler, Kicker design for the ELETTRA/SLS longitudinal multi-bunch feedback, Proc. EPAC2002, Paris, 2002.

[44] http://www.mathworks.com/

[45] H. Hindi *et al.,* Analysis of DSP-based longitudinal feedback system: Trials at SPEAR and ALS, Proc. PAC'93, Washington, 1993.

[46] H. Hindi *et al.,* A formal approach to the design of multibunch feedback systems: LQG controllers, Proc. EPAC'94, London, 1994.

[47] J. Fox *et al.,* Down sampled signal processing for a B Factory bunch-by-bunch feedback system, Proc. EPAC'92, Berlin, 1992.

[48] S. Prabhakar *et al.,* Calculation of impedance from multibunch synchronous phases: Theory and experimental results (SLAC-PUB-7979, 1998).

[49] D. Teytelmen *et al.,* Frequency resolved measurement of longitudinal impedances using transient beam diagnostics, Proc. PAC2001, Chicago, 2001.

[50] J. Fox *et al.,* Multi-bunch instability diagnostics via digital feedback systems at PEP-II, DAΦNE, ALS and SPEAR, Proc. PAC'99, New York, 1999.

[51] D. Teytelman *et al.,* Observation, control and modal analysis of longitudinal coupled-bunch instabilities in the ALS via a digital feedback system, (SLAC-PUB-7292, LBNL-39329 1996).

[52] S. Khan, Diagnostics of instabilities at BESSY II using the digital longitudinal feedback system, Proc. EPAC2002, Paris, 2002.

[53] A. Drago, Horizontal instability and feedback performance in DAΦNE E+ ring, Proc. EPAC2004, Lucerne, 2004.

**Bibliography**

H. Winick, *Synchrotron Radiation Sources – A Primer* (World Scientific, Singapore, 1994).

A.W. Chao, *Physics of Collective Beam Instabilities in High Energy Accelerators* (Wiley, New York, 1993).

A.V. Oppenheim and R.W. Schafer, *Digital Signal Processing* (Prentice-Hall, Englewood Cliffs, NJ, 1975).

G.F. Franklin, J.D. Powell and A. Emami-Naeini, *Feedback Control of Dynamic Systems* (Addison-Wesley, Reading, MA, 1994).

S. Prabhakar, New diagnostics and cures for coupled-bunch instabilities (PhD thesis, Stanford University, SLAC-Report-554, 2000).

D. Teytelman, Architectures and algorithms for control and diagnostics of coupled-bunch instabilities in circular accelerators (PhD thesis, Stanford University, SLAC-Report-633), 2003.